\documentclass[a4paper,onecolumn,11pt,accepted=2026-06-10]{quantumarticle}
\pdfoutput=1
\usepackage[utf8]{inputenc}
\usepackage[english]{babel}
\usepackage[T1]{fontenc}
\usepackage{amsmath}
\usepackage{amssymb}
\PassOptionsToPackage{hyphens}{url}\usepackage{hyperref}
\usepackage{cleveref}
\usepackage{xcolor,soul}
\crefname{equation}{Eq.}{Eqs.}
\Crefname{equation}{Equation}{Equations}
\crefname{figure}{Fig.}{Figs.}
\Crefname{figure}{Figure}{Figures}
\crefname{section}{Sect.}{Sects.}
\Crefname{section}{Section}{Sections}
\usepackage{parskip}

\usepackage{tikz}
\usepackage{lipsum}
\usetikzlibrary{quantikz2}

\usepackage{graphicx}
\newcommand{\cev}[1]{\reflectbox{\ensuremath{\vec{\reflectbox{\ensuremath{#1}}}}}}
\usepackage{extarrows}

\newcommand{\pp}[1]{\left({#1}\right)}
\newcommand{\bb}[1]{\left[{#1}\right]}
\usepackage{physics}
\usepackage{centernot}

\DeclareFontFamily{U}{stixextrai}{}
\DeclareFontShape{U}{stixextrai}{m}{n}
 { <-> stix-mathtt }{}

\newcommand{\lltriangle}{{\usefont{U}{stixextrai}{m}{n}\symbol{153}}}

\makeatletter
\providecommand{\vrectangle}{\mkern1mu\mathpalette\v@rectangle\relax\mkern1mu}
\newcommand{\v@rectangle}[2]{%
  \hbox{
  \fboxrule=0.5\fontdimen 8
    \ifx#1\displaystyle\textfont\else
    \ifx#1\textstyle\textfont\else
    \ifx#1\scriptstyle\scriptfont\else
    \scriptscriptfont\fi\fi\fi 3
  \fboxsep=-\fboxrule
  \fbox{$\m@th#1\phantom{(}$}%
  }
}
\makeatother

\usepackage{algorithm}
\usepackage{algorithmic}

\usepackage{tikz}
\usetikzlibrary{quantikz2}

\begin{document}
\title{Enlarging the GKP stabilizer group for enhanced noise protection}
\author{Jonathan Pelletier}
\author{Baptiste Royer}
\affiliation{Département de Physique, Université de Sherbrooke, Sherbrooke, Qc. J1K 2R1, Canada}
\affiliation{Institut Quantique, Université de Sherbrooke, Sherbrooke, Qc. J1K 2R1, Canada}
\maketitle
\begin{abstract}
    \noindent Encoding a qubit in a larger Hilbert space of an oscillator is an efficient way to protect its quantum information against decoherence. Promising examples of such bosonic encodings are the Gottesman-Kitaev-Preskill (GKP) codes. In this work, we investigate how redefining the stabilizer group of the GKP codes to include all operations with trivial action on the code space can contribute to the search for an optimal implementation of a logical circuit when it is affected by noise. We find the generators of the Gaussian stabilizer group, allowing us to search for different physical implementations of a Clifford operation. We then propose an algorithm that finds the optimal implementation of a given logical Clifford circuit on GKP codes, such that the state is less affected by loss errors during the computation. Finally, we demonstrate numerically, with logical randomized benchmarking, that such a compiler can increase the lifetime of square-GKP qubits while running Clifford circuits, compared to a random walk compiler. 
\end{abstract}
\section{Introduction}
As we move toward and beyond advanced noisy intermediate-scale quantum computing, the roles of quantum error correction (QEC) and compilation techniques become increasingly important. QEC ensures that the errors caused by decoherence channels during a computation do not accumulate too quickly, while compilation techniques can reduce the rate at which they occur in the first place. 
The combination of both is essential to move toward fault-tolerant computations and useful applications. QEC consist of redundantly encoding quantum information in a larger quantum system such that small physical errors can be detected and corrected before the full system is affected by its spreading. Systems that enable bosonic encodings are promising platforms for QEC, where the Hilbert space of possibly many high-quality harmonic oscillator modes can be used to encode quantum information. 
The increase in the ratio of redundancy over the number of physical devices needed compared to other options, like qubit codes, is an encouraging perspective toward the scaling of such implementations in a hardware-efficient manner. 
Bosonic codes can be implemented in many architectures, such as cavity quantum electrodynamics \cite{sivak_2023,lachance-quirion_2024,brock_2025,ofek_2016}, optical modes 
\cite{konno_2024,larsen_2025} and trapped ions \cite{deneeve_2022,matsos_2024a,matsos_2025}, to cite only a few. Amplitude damping (bosonic loss) is the main type of error channel affecting those architectures, followed by dephasing and other higher-order non-linear processes. 
The code introduced by Gottesman, Kitaev and Preskill (GKP)~\cite{gottesman_2001} can correct small displacement errors, and has been observed to be very efficient to protect against amplitude damping, while remaining competitive against other types of noise~\cite{noh_2019,leviant_2022}. 
This has created a lot of interest to study the limitations of GKP codes and potential ways to achieve universal quantum computation. For example, with the ability to apply arbitrary Clifford gates, it is possible to achieve universal computation with the addition of magic state injection \cite{bravyi_2005}. This path to universal computation is particularly interesting for GKP codes because Clifford operations can be realized with Gaussian operations only, for which the spread of displacement error can be controlled. If we instead consider finite-energy GKP codes and the bosonic loss, the effect of the error channel becomes dependent on the particular Gaussian operations used to implement the logical operation. Solving this issue and finding good compiling algorithms for Clifford circuits on finite energy GKP codes is then crucial to achieve universal quantum computing with bosonic codes.

\noindent In the present work, we study this problem of finding an optimal implementation of a Clifford circuit on a multimode finite energy GKP encoding, while it is affected by bosonic loss. To attain this goal, we first identify all logically equivalent implementations of a given logical operation on the ideal code. It is parametrized by the stabilizer group, which we define here to be the (potentially non-abelian) group of all operations that stabilize the code space. 
This leads us to include the standard translations symmetries of GKP codes, but also an infinite number of non-commuting stabilizers. Focusing on Clifford operations, we restrict this group to Gaussian operations and define the Gaussian GKP stabilizer group. Along the way, we solve a quadratic matrix equation over the restricted ring of integer matrices to fully characterize the generators of the group. Moving on to an application on finite energy GKP, we find that each GKP codeword implementation can be described by a general Fock-type envelope that generalizes the concept of finite-energy GKP codewords~\cite{royer_2020}. For the sake of completeness, we include an updated version of the so-called small-Big-small (sBs) stabilization protocol for those codewords. We find that the logical information contained within a finite energy GKP state is independent of the envelope, and is only dependent on the underlying logical grid, which create redundancy on how we choose to implement a logical state. A consequence of the redundancy is that choosing a specific implementation of a gate is equivalent to choosing an implementation of a codeword, which is also equivalent to choosing an envelope. This enables us to find clear optimization criteria on Clifford circuits based on the characteristics of the bosonic loss channel and their effects on the envelope of the codewords. In the last part of this work, we propose an algorithm that compiles Clifford circuits using those criteria. Finally, we use logical randomized benchmarking to compare our compiler algorithm to a random walk compiler algorithm \cite{tzitrin_2020} and show the gain of using the former. In the appendices, we included supplementary results that are more technical, but still of interest for the community. In \cref{app_sBs}, we present the generalization of the sBs protocol to any Fock-type envelope. In \cref{App_Wig_proj,App_Wig_FE_states,App_Wig_number}, we included a derivation of the Wigner function of the codespace projector for any GKP code, the derivation of the Wigner function of any finite energy GKP grid state and a computation of the number of excition contained in a finite energy GKP grid state, respectively.

\noindent The paper is organized as follows. We start in \cref{sec:pan} with useful definitions and notations on GKP codes, their representation and operations on harmonic oscillators. In \cref{sec:gsg}, we present and characterize the Gaussian stabilizer group for GKP codes. It is followed by its application on finite energy GKP codes to find optimization criteria in \cref{sec:cct}. In \cref{sec:gsc}, we introduce our compiler algorithm and analyze its performance. We conclude in \cref{sec:C}.

\section{\label{sec:pan}Preliminaries and notation}
In this section, we review the construction of multimode GKP codes as lattice codes \cite{gottesman_2001,harrington_2001,conrad_2022,royer_2022}. GKP codes are stabilizer codes which use the infinite-dimensional quantum Hilbert space of $N$ bosonic modes to encode a finite-dimensional logical subspace.
We denote the creation and annihilation operators for the $k^\text{th}$ mode as $\hat{a}_k$ and $\hat{a}^\dagger_k$, respectively, with commutation relation $[\hat{a}_j,\hat{a}^\dagger_k] = \delta_{j,k}$ for $j,k=1,...,N$. We also make use of the quadrature coordinates $\hat{q}_k = (\hat{a}_k+\hat{a}^\dagger_k)/\sqrt{2}$ and $\hat{p}_k = -i(\hat{a}_k-\hat{a}^\dagger_k)/\sqrt{2}$ with $\hbar=1$. We arrange these coordinates in a vector of operators $\mathbf{\hat{\boldsymbol{\xi}}} = (\mathbf{\hat{q}},\mathbf{\hat{p}})^\text{T}$ in the $qqpp$ convention
such that the commutation relation becomes $[\hat{\xi}_j,\hat{\xi}_k] = i \Omega_{jk}$ for $j,k=1,...,2N$. The matrix
\begin{equation}
  \Omega_{2N} =  \begin{pmatrix}
    0 & 1 \\
 -1 & 0 
    \end{pmatrix} \otimes I_N
\end{equation}
gives rise to the skew-symmetric bilinear form (symplectic form)
\begin{equation}
  \omega(\mathbf{u},\mathbf{v}) = \mathbf{u}^T\Omega \mathbf{v},
\end{equation}
defined for $\mathbf{u},\mathbf{v}\in \mathbb{R}^{2N}$. We omit the subscript of $\Omega_{2N}$ when the dimension is clear from the context. We define the translation
operator as 
\begin{equation}
 \hat{T}\left(\mathbf{v}\right) = {\rm e}^{-il \mathbf{v}^T \Omega\boldsymbol{\hat{\xi}} },
\end{equation}
from which the relation
\begin{equation}\label{eq:trans_relation}
 \hat{T}\pp{\mathbf{u}}\hat{T}\pp{\mathbf{v}} = {\rm e}^{-i l^2\mathbf{u}^T\Omega\mathbf{v}}\text{ } \hat{T}\pp{\mathbf{v}}\hat{T}
 \pp{\mathbf{u}} ={\rm e}^{\frac{-i l^2}{2}\mathbf{u}^T\Omega\mathbf{v}}\text{ } \hat{T}\pp{\mathbf{u}+\mathbf{v}}
\end{equation}
is implied by the commutation relation defined before. When using translations
with units of $l=\sqrt{2\pi}$, the commutativity of two translation operators is  
tantamount to the symplectic form being an integer:
\begin{equation}\label{commute_condition}
 [\hat{T}\pp{\mathbf{u}},\hat{T}\pp{\mathbf{v}}] = 0 \iff \omega\pp{
 \mathbf{u},\mathbf{v}} \in \mathbb{Z}. 
\end{equation}
These translation operators form representations of the Heisenberg group $H_{2N+1}\pp{\mathbb{R}}$
on the infinite Hilbert space of $N$ oscillators. GKP codes are stabilizer codes
which use this structure to embed one or multiple finite-dimensional representations
of the Weyl-Heisenberg group on a finite-dimensional logical subspace using the translation operators. In more detail, the stabilizer of a GKP code is an abelian group generated by a set of $2N$ translation operators
\begin{equation}
 \mathcal{S} = \left< {\rm e}^{2\pi i\mu_1}\hat{T}(\mathbf{s}_1),{\rm e}^{2\pi i\mu_2}\hat{T}(\mathbf{s}_2),...,{\rm e}^{2\pi i\mu_{2N}}\hat{T}(\mathbf{s}_{2N})\right>,
\end{equation}
with $\{\mathbf{s}_j\}$ a basis for $\mathbb{R}^{2N}$ and $\boldsymbol{\mu}$ the gauge vector. In most cases we take the trivial gauge $\boldsymbol{\mu}=0$. The abelianity condition of $\mathcal{S}$ translate
to the matrix
\begin{equation}
 A = S\Omega S^T
\end{equation} 
being integral, where the matrix $S$ is constructed from $\{\mathbf{s}_j\}$ as row vectors.
We refer to $A$ as the symplectic Gram matrix.
The GKP stabilizer group is then isomorphic to a lattice $\Lambda \subset \mathbb{R}^{2N}$ generated by the matrix $S$ with
the elements given by
\begin{equation}\label{eq:stab}
  \Lambda = \{S^T\mathbf{a}\text{ } | \text{ } \mathbf{a} \in \mathbb{Z}^{2N}\}.
\end{equation}
In the stabilizer formalism, we obtain the generalized logical Pauli operators as operators that commute with all of the stabilizers, but are not in the stabilizer:
$\mathcal{C}(\mathcal{S})/\mathcal{S}$. Using the isomorphism between the stabilizer group and the lattice $\Lambda$, we can show that the Pauli operators coincide with the definition of the symplectic dual lattice $\Lambda^*$
\begin{equation}
  \Lambda^{*} = \{\mathbf{{u}} \text{ } | \text{ } S\Omega\mathbf{u} \in \mathbb{Z}^{2N}\}.
\end{equation}
A potential generator matrix for the symplectic dual lattice is 
\begin{equation}
 S^{*} = A^{-1}S , 
\end{equation}
such that we can also write
\begin{equation}
  \Lambda^* = \{\Omega S^{-1}\mathbf{b}\text{ } | \text{ } \mathbf{b} \in \mathbb{Z}^{2N}\}.
\end{equation}
As a result, the logical Pauli operators are each associated with one of the $\text{det}(\Lambda) = \text{det}(A) = d^2$ cosets of $\Lambda^*/\Lambda$,
where $d$ is the dimension of the finite-dimensional encoded Hilbert space.

An important type of unitary operation on oscillators is Gaussian operations,
which are generated by Hamiltonians quadratic in mode operators
\begin{equation}\label{unitary_rep}
 \hat{U}_M = {\rm \exp}\left(\frac{-i}{2}\boldsymbol{\hat{\xi}}^T C \boldsymbol{\hat{\xi}} \right), \quad C=C^T,
\end{equation}
where $C$ is a symmetric matrix that specifies the generating Hamiltonian. The group of Gaussian unitaries can also be parametrized with symplectic matrices $M\in {\rm Sp}_{2N}(\mathbb{R})$, where the connection with \cref{unitary_rep} is $M = {\rm exp}(\Omega C)$. The unitaries $\hat{U}_M$ are representations of the metaplectic group, which is a double cover of the symplectic group. However, it is possible to lift those projective representations to ordinary representations by choosing the right representatives for each $M$. Doing so, the unitaries respect the homomorphism property
\begin{equation}
 \hat{U}_{N}\hat{U}_{M} = \hat{U}_{NM}.
\end{equation}
The action of Gaussian unitaries on mode operators is a linear application of
the symplectic representative
\begin{equation}
 U_M^\dagger \boldsymbol{\hat{\xi}} U_M = M\boldsymbol{\hat{\xi}}, 
\end{equation}
where we have an element-wise product on the left and matrix multiplication on the
right. This implies the more general relation
\begin{equation}\label{eq:quadrature_update}
 U_M^\dagger f\pp{\boldsymbol{\hat{\xi}}} U_M = f\pp{M\boldsymbol{\hat{\xi}}},
\end{equation}
for any function $f(\boldsymbol{\hat{\xi}})$ that can be represented
as a power series of mode operators. Any symplectic matrix $M$ can be written as a
a product of two orthogonal matrices and a positive definite diagonal matrix via its
Bloch-Messiah decomposition:
\begin{equation}
 M = O_2\Gamma O_1,
\end{equation}
where $O_i\in \mathrm{Sp}(2N,\mathbb{R}) \cap \mathrm{O}(2N) \cong U(N)$ and $\Gamma = D\oplus D^{-1}$
with $D$ is a positive definite diagonal matrix of dimension $N$. 

In this work, we simulate GKP states directly using their Wigner function, defined as
\begin{equation}\label{eq:wigner_def}
 W\pp{\boldsymbol{\xi},\hat{\rho}} = \int_{\mathbb{R}^{2N}}d\boldsymbol{v}\, \mathrm{e}^{-i{2\pi}\boldsymbol{\xi}^T\Omega \boldsymbol{v}}\, \text{Tr}\bb{\hat{T}\pp{\boldsymbol{v}}\hat{\rho}}.
\end{equation}
This version of the Wigner function is corrected for our specific version of the translation operator and the $l = \sqrt{2\pi}$ scaling for the units of $\boldsymbol{\xi}$. Consequently, the average of an operator $\hat{A}$ in phase-space is given by \begin{equation}
 \left<\hat{A}\right> = \text{Tr}\pp{\hat{\rho}\hat{A}} = \pp{2\pi}^{2N}\int_{\mathbb{R}^{2N}}d\boldsymbol{\xi} \, W\pp{\boldsymbol{\xi},\hat{\rho}}W\pp{\boldsymbol{\xi},\hat{A}}.
\end{equation}
When a general gaussian unitary acts on the density matrix $U_{M,\boldsymbol{\lambda}}\,:\, \hat{\rho}\to U_{M,\boldsymbol{\lambda}}\,\hat{\rho}\,U_{M,\boldsymbol{\lambda}}^\dagger$, the Wigner function is modified to
\begin{equation}\label{eq:mod_wigner}
 W\pp{\boldsymbol{\xi}, U_{M,\boldsymbol{\lambda}}\,\hat{\rho}\,U_{M,\boldsymbol{\lambda}}^\dagger} = W\pp{M^{-1}\bb{\boldsymbol{\xi}-\boldsymbol{\lambda}},\hat{\rho}},
\end{equation}
where $U_{M,\boldsymbol{\lambda}} = \hat{T}\pp{\boldsymbol{\lambda}}\hat{U}_M$. An important class of states in harmonic oscillators are Gaussian states. They are the states that are fully described by their first and second moments:
\begin{equation}
 \boldsymbol{\mu} = \left<\boldsymbol{\hat{\xi}}\right>/\sqrt{2\pi},\qquad \Sigma_{jk} = \frac{1}{4\pi}\left<\left\{ \boldsymbol{\hat{\xi}}_j-\sqrt{2\pi}\boldsymbol{\mu}_j,\boldsymbol{\hat{\xi}}_k-\sqrt{2\pi}\boldsymbol{\mu}_k  \right\}    \right>,
\end{equation}
where $\left\{\cdot,\cdot \right\}$ is the anticommutator. In phase-space, they are represented by gaussian Wigner functions 
\begin{equation}\label{eq:gauss_wig}
 G_{\Sigma,\boldsymbol{\mu}}\pp{\boldsymbol{\xi}} =  \frac{\exp\pp{-\frac{1}{2}\pp{\boldsymbol{\xi}-\boldsymbol{\mu}}^T\Sigma^{-1}\pp{\boldsymbol{\xi}-\boldsymbol{\mu}}}}{\sqrt{\pp{2\pi}^{2N}\text{det}\pp{\Sigma}}}.
\end{equation}
Finally, we define, on a bipartite qubit system $jk$, the generalized controlled Pauli operation of applying Pauli operator $B$ on subsystem $k$ if subsystem $j$ is in the $-1$ eigenstate of Pauli operator $A$ as 
\begin{equation}
 C\pp{A_j,B_k} = \frac{I_j I_k+A_j I_k+I_j B_k-A_j B_k}{2}.
\end{equation}
This symbol is symmetric in the sense that $C\pp{A_j,B_k}=C\pp{B_k,A_j}$. Using this notation, a controlled $Z$ gate is noted as $C\pp{Z_1,Z_2}$ and a controlled $X$ gate with control qubit 1 and target qubit 2 is noted as $C\pp{Z_1,X_2}$.

\section{\label{sec:gsg}Gaussian stabilizer group}
\subsection{Generalized Stabilizer Groups}
Qubit stabilizer codes use $n$ qubits to encode $k\leq n$ qubits. The codespace $\mathcal{T}$ of these codes is the submanifold of the full Hilbert space $\pp{\mathbb{C}^2}^n$ restricted by the elements of the stabilizer group $\mathcal{S}$ containing Pauli operators such that
\begin{equation}
\mathcal T = \left\{ \ket{\psi} : s\ket{\psi} = \ket{\psi} \forall s\in\mathcal{S}  \right\}.
\end{equation}
In the usual definition, the stabilizer group of a qubit code is generated by $n-k$ commuting elements, and it completely defines the code \cite{gottesman_1997}. In this definition of the stabilizer group, it is taken to be abelian. However, it is possible to relax this assumption of abelianity and only ask that the stabilizers need to share the $2^k$ eigenvectors that describe $T$ with eigenvalue $1$. This is equivalent to asking that the action of the stabilizers on the codespace commute, but not necessarily the operators themselves: 
\begin{equation}
 S_j S_k \ket{\psi} = S_k S_j\ket{\psi}\; \forall\; \ket{\psi}\in T \centernot\implies \bb{S_j,S_k} =0.
\end{equation}
The XS \cite{ni_2015} and XP \cite{webster_2022} formalisms are two examples where qubit codes were constructed from a non-abelian stabilizer group. In general, removing this restriction to abelian groups enables us to define a new stabilizer group that we call the unitary stabilizer :
\begin{equation}\label{eq:uni_stab}
 \mathcal{S}_{U} = \left\{ U \in U\pp{2^n} | \text{ } U\ket{\psi} = \ket{\psi} \forall \ket{\psi} \in T \right\}.
\end{equation}
The stabilizer group described in \cite{gottesman_1997} is, by definition, a subgroups of $\mathcal{S}_{U}$. Similar to the abelian elements, the unitary stabilizers can be used as a resource for stabilization, error mitigation, and potentially decoding. This notion of unitary stabilizer groups can also be used for continuous-variable codes such as the GKP encoding, in which case it becomes of infinite order. It is worth noting that, contrary to the abelian stabilizer, unitary stabilizers might mix error spaces or even entangle those subspaces together, which might be devastating for the preservation of information in the system. With the definition of \cref{eq:uni_stab}, we can understand those symmetry operations as encoded identities that are implemented via a non-identity operation on the system.

A way to identify such symmetry of a code is to compare the group order of a logical operation with its physical implementation. If they have a mismatching order, then necessarily there will exist a symmetry that will be included in $\mathcal{S}_U$. For example, let $g$ be a unitary operation on an arbitrary Hilbert space that implements a logical operator $\overline{g}$ of finite order $|\overline{g}|$. If $g$ is an adequate representation of $\overline{g}$, then, necessarily,
\begin{equation}
  g^{r|\overline{g}|} \in \mathcal{S}_U \quad \forall r\in \mathbb{Z},
\end{equation}  
as $g^{|\overline{g}|}$ must implement a logical identity. An example of such operations for the square GKP code is the logical gate $\sqrt{\,\overline{H}\,}$ implemented via the unitary \cite{royer_2022,boudreault_2025}:
\begin{equation}
 \text{Imp}\pp{\sqrt{\,\overline{H}\,}} = \exp\pp{\frac{i\pi}{8}\hat{n}^2}.
\end{equation}
Since $\hat{n}$ has integer eigenvalues, this physical operation has order $16$, 
while the underlying logical operation, $\sqrt{\, \overline{H}\,}$, has order $4$. As a result, the four unitaries
\begin{equation}
 \left\{  \exp\pp{\frac{ir\pi}{2}\hat{n}^2} \right\}_{r\in\mathbb{Z}_{4}},
\end{equation}
implement a logical identity on the square GKP code. For infinite-dimensional Hilbert spaces, the physical implementation of a logical operation might be of infinite order, resulting in the inclusion of an infinite subgroup in $\mathcal{S}_U$. This case will arise in our treatment of the GKP code with squeezing operations. There might even exist stabilizing operations that do not correspond to powers of non-trivial logical operations. As a whole, the rich structure of the unitary stabilizer group makes it impractical to try to find every element in $\mathcal{S}_U$. Following this thread of ideas, we restrict the scope of this paper to grid codes. We will also only be interested in a subgroup that is more easily implementable, \textit{i.e.} Gaussian operations. For qubit stabilizer codes, the restriction of the unitary stabilizer group to Clifford operations has been studied \cite{rengaswamy_2020,kuehnke_2025} to optimize the implementation of logical Clifford operations. This work extends this notion to GKP codes.  
\subsection{Gaussian Stabilizer Group}\label{GSG}
The elements of any generalized stabilizer group on grid codes need to share a specific property: they need to stabilize the Pauli subgrids $\mathcal{P}_j \subset \Lambda^*$ individually:
\begin{equation}
 \mathcal{P}_j = \Lambda + \mathbf{p}_j\quad \mathbf{p}_j\in\Lambda^*,
\end{equation}
with an index $j$ for every logical Pauli string. By definition, $\mathcal{P}_0$ is $\Lambda$. This gives a new definition of the group as the intersection of all automorphisms of these subgrids
\begin{equation}\label{SU_AUT_star}
 \mathcal{S}_U = \bigcap_{j}\text{Aut}\pp{\mathcal{P}_j} \subset \text{Aut}\pp{{\Lambda}^*} .
\end{equation}
For grid codes, an important subgroup of $\mathcal{S}_U$ is the Gaussian stabilizer group, which we define as the intersection of the unitary stabilizer group and the Gaussian unitaries
\begin{equation}
 \mathcal{S}_{G} = \mathcal{S}_U \cap \left\{ U_M| M\in{\rm Sp}_{2N}(\mathbb{R})\right\}.
\end{equation}
Since any Gaussian unitary acts as an affine transformation on phase space, and stabilizers in $\mathcal{S}_G$ must be logical operations of the code, we know that each element of $\mathcal{S}_G$ admits the decomposition \cite{conrad_2024}
\begin{equation}
 S_G \subset \text{Aut}_{\infty}^{S}\pp{{\Lambda}^*} = \text{Aut}^{S}\pp{{\Lambda}^*} \ltimes {\Lambda}^*.
\end{equation}
Such affine transformations are characterized by two transformations: a fixed point symplectic transformation $M\in \text{Aut}^{S}\pp{{\Lambda}^*}$ and a translation by a vector of $\boldsymbol{\lambda}^*\in{\Lambda}^*$. We denote such transformations by $\mathcal{L}_{M,\lambda^*}$ with its application
\begin{equation}
 \mathcal{L}_{M,\boldsymbol{\lambda}^*}\pp{\boldsymbol{u}} = M\boldsymbol{u} + \boldsymbol{\lambda}^*.
\end{equation}
With respect to the condition in \cref{SU_AUT_star}, these morphisms describe an element of $S_G$ when
\begin{equation}
 \mathcal{L}_{M,\boldsymbol{\lambda}^*}\pp{\mathcal{P}_j} = \mathcal{P}_j \quad \forall j. 
\end{equation}
Expanding the restrictriction in terms of the original lattice $\mathcal{P}_0$ moved by a vector of the dual lattice $\boldsymbol{p}_j$, we obtain a much simpler condition on $M$:
\begin{equation}
 \mathcal{L}_{M,\boldsymbol{\lambda}^*}\pp{\mathcal{P}_j} = \mathcal{L}_{M,\boldsymbol{\lambda}^*}\pp{\mathcal{P}_0} + M\mathbf{p}_j = \mathcal{P}_0 + \mathbf{p}_j,
\end{equation}
\begin{equation}\label{iso_eq}
 \pp{M-I}\mathbf{p}_j \in\Lambda \quad \forall j.
\end{equation}
Using both the lattice and the dual lattice definition in terms of their generator matrix ($S$ and $S^*$), we conclude that elements of $S_G$ must have a fixed-point transformation of the form
\begin{equation}\label{symp_stab_def}
 M = S^T \pp{XA + I} S^{-T},
\end{equation}
where $X\in \text{Mat}_{2N}\pp{\mathbb{Z}}$ is a general matrix of integers and $\pp{XA + I}\in \text{SL}_{2N}\pp{\mathbb{Z}}$ (see \cref{App_Symp_demo}). The matrix $M$ is a valid solution only if it is also symplectic, which is not implied by \cref{symp_stab_def}. Enforcing the symplectic structure to $M$ returns an equation for $X$
\begin{equation}\label{XA_equation}
 XAX^T - \pp{X-X^T} = 0,
\end{equation}
with $A$ the symplectic Gram matrix that defines the lattice. From \cite{gottesman_2001,lin_2023}, we know that for every integrally symplectic matrix $A=S\Omega S^T$, there is a change of basis $R\in \text{SL}_{2N}\pp{\mathbb{Z}}$ that transforms $S$ and $A$ as
\begin{equation}\label{eq:canonical}
 S_D = RS, \quad A_D = RAR^T = \Omega_2 \otimes D,
\end{equation}
such that $D$ is a diagonal matrix of natural numbers that fix the type of lattice. Since $A$ and $A_D$ describe the same lattice $\Lambda$, the transformation $M$ must be valid for both. As a result, if the pair $\pp{X_D, A_D}$ is a solution to \cref{XA_equation}, then so must be $\pp{R^TX_DR, A}$. As a result, we only have to solve the equation
\begin{equation}\label{XAD_equation}
 X_DA_DX_D^T - \pp{X_D-X_D^T} = 0
\end{equation}
which is simpler since $A_D$ is skew-symmetric with only $N$ independent non-zero elements. The solutions of \cref{XAD_equation} form a group with a product $\circ$ defined as 
\begin{equation}\label{product_rule}
 X_D\circ Y_D = X_D + Y_D + X_DA_DY_D,   
\end{equation}
with $X_D,Y_D$ solutions to \cref{XAD_equation}. The group product for solutions in \cref{product_rule} is also the group product for solutions in a non-canonical basis, such as the solutions of \cref{XA_equation}.

\noindent To fully describe the transformations $M$, we now proceed to find the generators of this group of solutions using the generators of the symplectic algebra $\mathfrak{sp}\pp{2N, \mathbb{R}}$.
Looking at \cref{symp_stab_def}, we can deduce that $XA$ must be similar in structure to an element that is a small deviation from the identity, which also respects the symplectic property. This is equivalent to saying that $XA$ has the same structure as an element of the symplectic algebra $\mathfrak{sp}\pp{2N, \mathbb{R}}$. A basis for $\mathfrak{sp}\pp{2N, \mathbb{R}}$ can be chosen to be the union of two sets. The first set contains symmetric matrices
\begin{equation}
 F_{j,k} = E_{j,k} + E_{k,j}, \qquad F_{j,j} = E_{j,j},
\end{equation}
and the second contains skew-symmetric matrices
\begin{equation}
 G_{j,k} = E_{j,k} - E_{k,j},
\end{equation}
with $E_{j,k}$ the matrix with $1$ at position $(j,k)$ and $0$ elsewhere. We use those basis elements to search for solutions of \cref{XAD_equation}. Starting with the symmetric set, we fix the solution to $X_D = \alpha F_{j,k}$ with $k\geq j$ and try to solve for $ \alpha\in\mathbb{Z}$. This type of matrix is a non-zero solution of \cref{XAD_equation} only if 
\begin{equation}
 A_{D,j,k} = 0. 
\end{equation} 
This condition is validated for all values of $\alpha$ and all pairs of indices $\pp{j,k}$, except for combinations of the form $\pp{j,k} = \pp{j,N+j}$. We also find that the product rule in \cref{product_rule} is simplified to 
\begin{equation}
  \alpha_1 F_{j,k} \circ \alpha_2 F_{j,k} = \pp{\alpha_1+\alpha_2} F_{j,k}
\end{equation}
when restricted to this infinite subgroup of solutions. This means that solutions $F_{j,k}$ generate all other solutions of the form $\alpha F_{j,k}, \, \alpha\in\mathbb{Z}$ via products with itself. By counting the number of valid $(j,k)$ pairs, we obtain $2N^2$ generators of this type. Moving on to skew-symmetric solutions, we fix $X_D = \beta G_{j,k}$ with $\beta \in \mathbb{Z}$. This matrix is a solution only if 
\begin{equation}\label{involution}
  2-\beta A_{D,j,k} = 0.
\end{equation}
Because of the structure of $A_D$, this can only be validated for pairs of indices such that $\pp{j,k} = \pp{j,N+j}$, which complement the symmetric set of solutions. Additionally, the relation of \cref{involution} can also only be validated if the matrix $A_D$ contains elements $A_{D,j,k} \in\{1,2\}$. As such, we get two types of potential solutions: $\pp{\beta=1, A_{D,j,k}=2 }$ and $\pp{\beta = 2, A_{D,j,k}=1}$. Both correspond to an involution operation such that
\begin{gather}
  \beta G_{j,k} \circ \beta G_{j,k} = 0.
\end{gather}
Futhermore, only when $A_{D,j,k}=1$, we get an order four solution: $X_D = G_{j,k} + F_{j,j} + F_{k,k}$ that squares to the solution of \cref{involution}. In total, we have between $2N^2$ and $2N^2+N$ generators of solutions depending on the number of diagonal elements of $D$ that are either $1$ or $2$. This corresponds to the maximum number of generators possible, which means that we found all solutions to \cref{XAD_equation}.

\noindent Now that the fixed-point part of the transformation $M$ is determined, it only remains to fix $\boldsymbol{\lambda}^*$ to fully parametrize the transformation $\mathcal{L}_{M,\boldsymbol{\lambda}^*}$. The fixed-point transformation was obtained by fixing each Pauli grid to itself. However, this restriction does not restrain the phases that a Pauli representative can gain when the symplectic unitary is applied. This means that the symplectic unitary $U_M$ is, in the general case, an implementation of a logical Pauli operator. To adjust this phase, we choose $\boldsymbol{\lambda}^*$ such that the whole transformation commutes with every Pauli representative $\mathbf{p}_j$ on the code space
\begin{equation}\label{eq:rest}
  \hat{T}\left(\boldsymbol{\lambda}^*\right)U_M \hat{T}\left(\mathbf{p}_j\right)\ket{\psi} = \hat{T}\left(\mathbf{p}_j\right) \hat{T}\left(\boldsymbol{\lambda}^*\right)U_M \ket{\psi}.
\end{equation}
The Pauli representatives $\hat{T}\left(\mathbf{p}_j\right)$ might not commute in general with $\mathcal{L}_{M, \boldsymbol{\lambda}^*}$ as operators, but their action need to commute on the code space. Requiring that \cref{eq:rest} is true entails that the condition is also true after a transformation $\mathbf{p}_j\to \mathbf{p}_j + \boldsymbol{\lambda}\,\forall\,\boldsymbol{\lambda}\in\Lambda$. \Cref{eq:rest} leads to a set of $2N$ equations derived in appendix B that always have a solution $\boldsymbol{\lambda}^*\in\Lambda^*$. This choice fixes $\mathcal{L}_{M,\boldsymbol{\lambda}^*}$ up to a translation stabilizer of the code. In general, we only need to identify the action of $U_M$ on the Pauli representative and inverse it with a translation by the corresponding $\boldsymbol{\lambda}^*$.

\noindent In the end, for each generator of solution $X_k$ of \cref{XA_equation}, we get a gaussian stabilizer group generator $U_{M_k,\boldsymbol{\lambda_k}^*}$, that implements the transformations $\mathcal{L}_{M_k,\boldsymbol{\lambda_k}^*}$. We can now write the Gaussian stabilizer group as the group generated by the $2N$ translation stabilizers (indexed by $j$) and the $2N^2+m$ Gaussian stabilizer generators $U_{M_k,\boldsymbol{\lambda_k}^*}$
\begin{equation}\label{eq:gsg}
 \mathcal{S}_{G} = \left< \hat{T}(\mathbf{s}_j), U_{M_k,\boldsymbol{\lambda_k}^*}\right>
\end{equation} 
with $m$ the number of entries in $D$ that are either $1$ or $2$.
\subsection{Square GKP Gaussians Stabilizer}\label{squareGSG}
We now proceed to give an example of this group for the single-mode square GKP code and how it is related to the encoded Clifford group. We then construct the same group but for the combination of two square GKP codes before generalizing to $N$ square GKP codes. For the single mode case, the generator matrix is $S = \sqrt{2}I_2$ and the symplectic Gram matrix is
\begin{equation}
 A = 2\Omega_2,
\end{equation}
which is already in the canonical form. To parametrize the Gaussian stabilizer group, we are looking for all symplectic matrices of the form
\begin{equation}
 M = 2X\Omega_2+I.
\end{equation}
Following the analysis of the last section, we find three generators for the solutions of the \cref{XA_equation}. They are
\begin{equation}
 X_{H^2} = \begin{bmatrix}
    0 & 1 \\
 -1 & 0 
    \end{bmatrix}, \quad X_{Q^2} = \begin{bmatrix}
      1 & 0 \\
      0 & 0 
      \end{bmatrix}, \quad X_{P^2} = \begin{bmatrix}
        0 & 0 \\
        0 & 1 
        \end{bmatrix},
\end{equation}
with the corresponding symplectic matrices
\begin{equation}\label{Symp_sqrt}
 M_{H^2} = \begin{bmatrix}
 -1 & 0 \\
    0 & -1 
    \end{bmatrix}, \quad M_{Q^2} = \begin{bmatrix}
      1 & 2 \\
      0 & 1 
      \end{bmatrix}, \quad M_{P^2} = \begin{bmatrix}
        1 & 0 \\
 -2 & 1 
        \end{bmatrix}.
\end{equation}
We indexed them as $H^2$, $Q^2$ and $P^2$ because of their logical action on the code, with $Q$ denoting the phase gate in the logical $X$ basis and $P$ the phase gate in the logical $Z$ basis. The transformations in \cref{Symp_sqrt} implement the square of the logical operators $\overline{H}$, $\overline{\sqrt{X}}$ and $\overline{\sqrt{Z}}$, respectively. From each symplectic matrix of \cref{Symp_sqrt}, the corresponding unitary operator that implement it are obtained by solving \cref{unitary_rep}, which in this case yield ${\overline{H}^2} = \exp{-i \pi\hat{n}}$, ${\overline{Q}^2} = \exp{-i \hat{p}^2}$ and ${\overline{P}^2} = \exp{-i \hat{q}^2}$, respectively. It now only remains to fix the translation part of the affine transformations. Because the unitary $\overline{Q}^2$ has the same action on the logical Pauli representatives as an $\overline{X}$ operation (which can be obtained by solving the equations of \cref{app:restriction}), the affine transformation $\mathcal{L}_{M_{Q^2},\boldsymbol{\lambda}^*\in\mathcal{P}_x} = \overline{X}\,\overline{Q}^2$ is a generator of the gaussian stabilizer group. Similary, the transformation $\mathcal{L}_{M_{P^2},\boldsymbol{\lambda}^*\in\mathcal{P}_z} = \overline{Z}\, \overline{P}^2$ is also a generator. Finally, we find that $\overline{H}^2$ is itself a generator. The Gaussian stabilizer group of the single-mode square GKP code is
\begin{equation}\label{eq:gen_sq}
 \mathcal{S}_{\square} = \left< \overline{X}^2, \overline{Z}^2,\overline{X}\,\overline{Q}^2, \overline{Z}\,\overline{P}^2, \overline{H}^2\right>.
\end{equation} 
We now move on to the combination of two square GKP codes. In this case, the Gaussian stabilizer group inherit two copies of the Gaussian stabilizer group described previously, one for each mode. There are still four new generators to find out of the ten in total. The symplectic Gram matrix now is $2\Omega_4$, and we are looking for symplectic matrices of the form
\begin{equation}
 M = 2X\Omega_4+I.
\end{equation}
The four missing generators are related to the solutions $X = F_{j,k},\: k\geq j$ that are not on the diagonal of any single block, given by $F_{1,2},F_{2,3}, F_{3,4}$ and $F_{1,4}$. From those four solutions, we obtain the following four symplectic matrices
\begin{equation}
 M_1 = \begin{bmatrix}
    1 & 0 & 0 & 2 \\
    0 & 1 & 2 & 0 \\
    0 & 0 & 1 & 0 \\
    0 & 0 & 0 & 1 
    \end{bmatrix},\quad M_2 = \begin{bmatrix}
      1 & 0 & 0 & 0 \\
 -2 & 1 & 0 & 0 \\
      0 & 0 & 1 & 2 \\
      0 & 0 & 0 & 1 
      \end{bmatrix},
\end{equation}
\begin{equation}
 M_3 = \begin{bmatrix}
    1 & 0 & 0 & 0 \\
    0 & 1 & 0 & 0 \\
    0 & -2 & 1 & 0 \\
 -2 & 0 & 0 & 1 
    \end{bmatrix},\quad \text{and} \quad M_4 = \begin{bmatrix}
      1 & -2 & 0 & 0 \\
      0 & 1 & 0 & 0 \\
      0 & 0 & 1 & 0 \\
      0 & 0 & 2 & 1 
      \end{bmatrix}. 
\end{equation}
Their action on the Pauli operators $\overline{X}_j = \exp\pp{-i\hat{p}_j\sqrt{\pi}}$ and $\overline{Z}_j = \exp\pp{i\hat{q}_j\sqrt{\pi}}$, are given by the squares of the encoded controlled Pauli operators $\overline{C\pp{X_1,X_2}}$, $\overline{C\pp{Z_1,X_2}}$, $\overline{C\pp{Z_1,Z_2}}$ and $\overline{C\pp{X_1,Z_2}}$, in order from $M_1$ to $M_4$, respectively. The action of $\overline{C\pp{A,B}}$ is to apply a $\overline{B}$ operation if the state is in a $-1$ eigenstate of $\overline{A}$. The square of those controlled encoded gates are implemented by the unitaries
\begin{equation}
 \overline{C\pp{X_j,X_k}}^2 = \exp\pp{2i\hat{p}_j\hat{p}_k} \quad \overline{C\pp{Z_j,X_k}}^2 = \exp\pp{2i\hat{q}_j\hat{p}_k} \quad  \overline{C\pp{Z_j,Z_k}}^2 = \exp\pp{-2i\hat{q}_j\hat{q}_k}.
\end{equation}
Any encoded generalized controlled Pauli gate is an order two logical gate, such that $\overline{C\pp{A,B}}^2$ is in the Gaussian stabilizer group. As a result, we can write the full Gaussian stabilizer group as generated by the two sets of single-mode GKP Gaussian stabilizer generators and those four additional two-mode generators. Since the full group of symplectic operations is generated by two-body interactions, the Gaussian stabilizer group over $N$ single-mode square GKP is found as
\begin{equation}
 \mathcal{S}_{\square^{\otimes N}} = \left< \overline{X}_j^2, \overline{Z}_j^2,\overline{X}_j\,\overline{Q}_j^2, \overline{Z}_j\,\overline{P}_j^2, \overline{H}_j^2, \overline{C\pp{X_j,X_k}}^2,\: \overline{C\pp{Z_j,X_k}}^2, \overline{C\pp{Z_j,Z_k}}^2 \right>
\end{equation}
with $j,k\in \left\{1...N\right\}$ and $j\neq k$. Moreover, since this decomposition is based on the canonical form of the Gram matrix $A$, any combination of arbitrary one-mode GKP qubits leads to a Gaussian stabilizer group with the same structure and a very similar generator decomposition.

\section{\label{sec:cct} Noise protection of finite energy GKP codes}
The construction of the Gaussian stabilizer group in \cref{GSG,squareGSG} was tailored for GKP codes that span infinitely in phase space. In the following sections, we are interested in using this group as a resource to increase the performance of gate-based computation with a finite-energy GKP encoding. We show how a non-abelian stabilizer group on the infinite energy code becomes a logical isometry group on finite energy codewords. Based on arguments related to the noise channels, we propose optimization metrics to compare different implementations of Clifford gates.
\subsection{Finite-energy GKP and Gaussian stabilizer}\label{subsec:FEGKP}
\begin{figure}[t]
    \centering
    \includegraphics[scale=0.12]{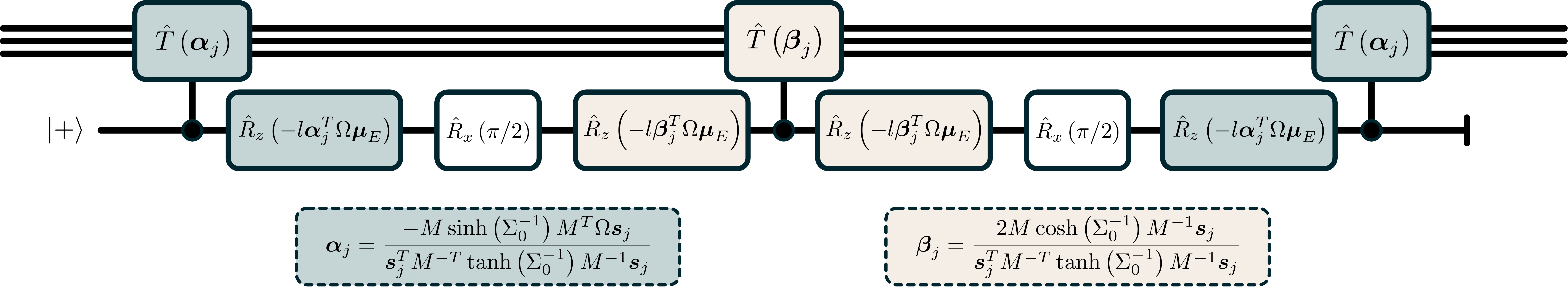}
  \caption{\textbf{Finite-energy GKP state stabilization.} Autonomous sBs stabilization circuit for a GKP state with envelope described by the covariance matrix $\Sigma_E = M\Sigma_0 M^T$ and position $\boldsymbol{\mu}_E$. The protocol uses engineered cooling of a GKP on $N$ modes (represented by three wires) using a 2-level system as an ancilla (bottom wire) that is discarded at the end. The full stabilization is the result of cycling over all the stabilizer generators $\left\{\boldsymbol{s}_j\right\}$ of the code.}
    \label{fig:sBs_circ}
\end{figure}

A logical state of the infinite energy GKP code defined by the generator matrix $S$ is noted $\ket{\Psi_{S}}$. We construct a finite-energy logical state from $\ket{\Psi_{S}}$ by acting with a fock type damping operator $E$ on it:
\begin{equation}\label{eq:fe_GKP_state}
 E\pp{\Sigma_E,\mu_E}\ket{\Psi_{S}} \propto \exp\pp{-\frac{1}{2}\bb{\boldsymbol{\hat{\xi}}-\boldsymbol{\mu}_E}^T \Sigma_E^{-1} \bb{\boldsymbol{\hat{\xi}}-\boldsymbol{\mu}_E}  }\ket{\Psi_{S}}.
\end{equation}
This damping operator can be understood as a Gaussian envelope described by a mean position $\boldsymbol{\mu}_E$ and a covariance matrix $\Sigma_E$. This state can be stabilized by a sequence of measurements of all normalized stabilizers
\begin{equation}\label{eq:norm_stab}
 \hat{T}_{\Sigma_E,\mu_E}\left(\mathbf{s}_j\right) = E\pp{\Sigma_E,\mu_E} \hat{T}\left(\mathbf{s}_j\right) E\pp{\Sigma_E,\mu_E}^{-1}. 
\end{equation}
Another way of achieving stabilization of this state is by engineering an oscillator-bath interaction that cools down the oscillator's system toward the $+1$ eigenspace of each $\hat{T}_{\Sigma_E,\mu_E}\left(\mathbf{s}_j\right) $. One such protocol, called sBs~\cite{deneeve_2022,royer_2020}, was described in Ref.~\cite{royer_2022} for a multimode GKP code with an isometric envelope. The sBs protocol can be generalized to accommodate a more general envelope $E\pp{\Sigma_E,\mu_E}$ presented here. The circuit that implements it, using only controlled displacements and qubit rotations, is presented in \cref{fig:sBs_circ} and the full derivation is included in \cref{app_sBs}. 
In an experimental setup, an initialization algorithm that prepares a logical state $\ket{\Psi_{S,\Sigma,\mu}} \propto E\pp{\Sigma_E,\mu_E}\ket{\Psi_{S}}$ with an envelope that is centred and distributed equally in the quadratures of each mode is preferred, as we show in the following sections. In other words, at the beginning of a computation, the ideal envelope is described by the parameters
\begin{equation}\label{eq:start_env}
   \Sigma_0 = \varepsilon^{-1} \oplus \varepsilon^{-1}\qquad \boldsymbol{\mu}_E=\boldsymbol{0} 
\end{equation}
where $\varepsilon$ is a diagonal matrix of dimension $N\times N$ with elements $\varepsilon_j$ on its diagonal that describe the distribution of mode $j$. At initialization time, the envelope thus takes the form
\begin{equation}\label{eq:init_env}
 E\pp{\Sigma_0,\boldsymbol{0}} = \exp\pp{-\sum_j \varepsilon_j \hat{n}_j}.
\end{equation}
The smaller the value of $\varepsilon_j$ is, the closer the code is to its infinite energy counterpart. In the process of stabilizing the GKP codespace in an actual device, there exists an optimal value $\varepsilon_j$, limited by the amount of amplitude damping or other noises.
For a unitary $U$ that implements a general logical operator on the infinite energy codespace $U\ket{\Psi_{S}} = \ket{\Phi_{S}}$, the application of the same operator on a finite-energy state is equivalent to applying the gate perfectly on the logical state and modifying the envelope:
\begin{equation}
 U E\pp{\Sigma,\boldsymbol{\mu}} \ket{\Psi_{S}} = \pp{U E\pp{\Sigma,\boldsymbol{\mu}} U^\dagger}U\ket{\Psi_{S}} = E^\prime\pp{\Sigma,\boldsymbol{\mu},U} \ket{\Phi_{S}}.
\end{equation}
For a general unitary $U$, the new envelope $E^\prime$ is not Gaussian. However, in the case that $U$ is a Gaussian unitary indexed by the symplectic matrix $M$ followed by a translation in phase space by $\boldsymbol{\lambda}$, $U_{M,\boldsymbol{\lambda}}$, the new envelope $E^\prime$ remains Gaussian and is computed using \cref{eq:quadrature_update}:
\begin{equation}\label{eq:Conj_trans}
 U_{M,\boldsymbol{\lambda}}  : E\pp{\Sigma,\boldsymbol{\mu} } \to E\pp{M\Sigma M^T,M\boldsymbol{\mu} + \boldsymbol{\lambda} }.
\end{equation}
As a result, no logical information is transferred to the envelope during a Clifford circuit computation since Clifford operations are implemented via Gaussian unitaries, . At each step, the logical information is contained only within the underlying infinite energy grid state $\ket{\Psi_{S}}$ or $\ket{\Phi_{S}}$. If the transformation $U_{M,\boldsymbol{\lambda}} $ describes a symmetry of the code, $U_{M,\boldsymbol{\lambda}} \ket{\Psi_{S}} = \ket{\Psi_{S}}$, the initial and final states both contain the same logical information, even if different envelopes describe the states. They are physically different, but logically equivalent. We've seen how to parametrize this set of logically equivalent states in \cref{sec:gsg}. The set of Gaussian operations that are also isometries of a GKP code is the definition of the Gaussian stabilizer group of \cref{eq:gsg}:
\begin{equation}
 U_{M,\boldsymbol{\lambda}} E\pp{\Sigma,\boldsymbol{\mu}}\ket{\Psi_{S}} = E\pp{M\Sigma M^T,M\boldsymbol{\mu}+\boldsymbol{\lambda}}\ket{\Psi_{S}}\quad \forall U_{M,\boldsymbol{\lambda}}\in \mathcal{S}_G.
\end{equation}
In the following, we show that the existence of an optimal state, and therefore, of an optimal circuit, is highly dependent on the noise model. 

\subsection{Noise model and state representative optimization}\label{seq:Noise_Mod}
A quantum circuit defined by a sequence of gates $\left\{V_k\right\}$ and initial logical state $\ket{{\Psi}_{0}}$ can, equivalently, be fully caracterised by a sequence $\left\{\ket{{\Psi}_{k}}\right\}$ of states : 
\begin{equation}
 ... \,\xrightarrow{{V}_{k-1}}\,\ket{{\Psi}_{k-1}}\, \xrightarrow{{V}_k}\,\ket{{\Psi}_{k}}\,\xrightarrow{{V}_{k+1}} \,...
\end{equation}
This dual description of a circuit can be used as a tool to optimize how the sequence $\left\{\overline{V}_k\right\}$ is implemented, based on the properties of the state sequence. During a computation, some intermediate states might be more prone to errors, making it favourable to choose an alternative circuit implementation and hence reducing the probability of having an error. In the context of quantum error correction, we aim to implement an encoded circuit $\left(\ket{\overline{\Psi}_{0}}, \left\{\overline{V}_k\right\} \right)$ onto a physical system :
\begin{equation}
 ... \,\xrightarrow{\text{Imp}\pp{\overline{V}_{k-1}}}\,\ket{\text{Imp}\pp{\overline{\Psi}_{k-1}}}\, \xrightarrow{\text{Imp}\pp{\overline{V}_k}}\,\ket{\text{Imp}\pp{\overline{\Psi}_{k}}}\,\xrightarrow{\text{Imp}\pp{\overline{V}_{k+1}}} \,...
\end{equation}
An explicit physical implementation (Imp) is called a compilation strategy. In the typical quantum error correction code, the implementation of a codeword $\text{Imp}\pp{\overline{\Psi}_{k}}$ is unique, and so, only the gate compilation needs to be optimized. As presented in the last section, the case of GKP codes is different since different physical states with different envelopes describe the same logical information. This additional degree of freedom can be used to reduce the effect of noise on the logical information during the computation by choosing the logical states that are less affected by noise at each step.
The optimization over all the different implementations can then be done in both the gates compilation picture or the states compilation picture. Since the best implementation is noise-dependent, we need to define a noise model and its effect on the GKP codewords as a function of the parameters of the envelope $\Sigma_E$ and $\boldsymbol{\mu}_E$ in order to find an optimization criterion.

\begin{figure}[t]
    \centering
    \includegraphics[width =\linewidth]{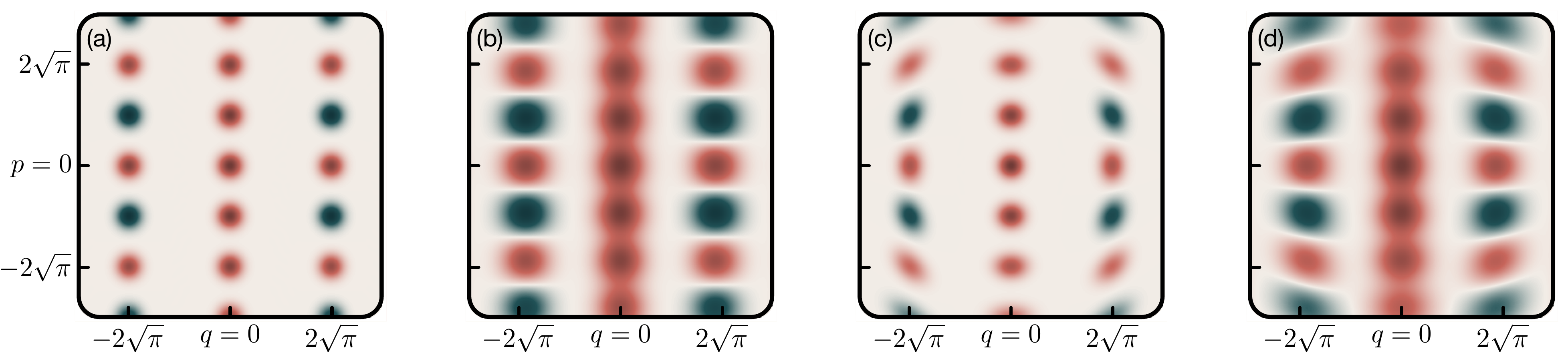}
  \caption{\textbf{Noise application on GKP states.} Effects of different error channels on the Wigner representation of the $\ket{\overline{0}}$ state of the square GKP code. (a) Unaffected state. (b) Amplitude damping with an average radial translation of $r\,\sqrt{\gamma}$. (c) Dephasing with an average absolute angular translation of $\sqrt{2/\pi}\,r\,\gamma_\phi$. (d) Combined amplitude damping and dephasing.}
    \label{fig:noise}
\end{figure}

\noindent There are two main error channels for bosonic codes in harmonic modes. The more prominent one is bosonic loss $\mathcal{N}_{L}\bb{\gamma}$ with its Kraus representation \cite{weedbrook_2012}:
\begin{equation}
 \mathcal{N}_{L}\bb{\gamma}\pp{\rho} = \sum_{k=0}^{\infty} \hat{L}_k\hat{\rho} \hat{L}_k^\dagger, \qquad \hat{L}_k = \sqrt{\frac{\gamma^k}{k!}}\pp{1-\gamma}^{\hat{n}/2}\hat{a}^k.
\end{equation}  
Among other transformations, the main action of this channel on functions of phase-space is a radial contraction of every feature toward the origin by a factor of $\sqrt{\gamma}$, displacing by an amount $\sqrt{\gamma}\,r$ features at radius $r$. The dephasing channel is the second most impactful noise channel on bosonic modes, and is represented as \cite{leviant_2022}
\begin{equation}\label{eq:Depha}
 \mathcal{N}_{D}\bb{\gamma_\phi}\pp{\rho} = \frac{1}{\sqrt{2\pi\gamma_\phi}}\int_{\mathbb{R}}\mathrm{d}\phi\, e^{-\phi^2/2\gamma_\phi} e^{i\phi\hat{n}}\hat{\rho} e^{-i\phi\hat{n}}.
\end{equation}
Its action is to smear phase-space functions in the angular direction, which can also be interpreted as a rotation in phase-space by a random angle sampled from a normal distribution. On average, this channel rotates the Wigner function of a state by an angle of amplitude $\sqrt{2/\pi}\,\gamma_\phi$. For a small dephasing rate $\gamma_\phi$, this induces a displacement of $\sqrt{2/\pi}\,\gamma_\phi \, r$ for a feature positioned at radius $r$ in phase-space. Because their action are in orthogonal directions, the pure loss and dephasing channel commute \cite{weedbrook_2012} and, therefore, the general noise channel can be described as a composition of both:
\begin{equation}
 \mathcal{N}_{LD}\bb{\gamma,\gamma_\phi} = \mathcal{N}_{L}\bb{\gamma}\circ \mathcal{N}_{D}\bb{\gamma_\phi}=  \mathcal{N}_{D}\bb{\gamma_\phi}\circ \mathcal{N}_{L}\bb{\gamma}.
\end{equation} 
When both channels are active, the amplitude of the total translation of a feature at radius $r$ is, on average,
\begin{equation}
  \Delta\boldsymbol{\xi}\pp{r} = r\,\sqrt{\gamma+\frac{2}{\pi}\gamma_\phi^2}.
\end{equation}
Coming back to finite energy GKP codes, each feature of the Wigner function of a GKP state is a Gaussian (more details in \cref{Perf_GSC}). When both bosonic loss and dephasing are considered, each Gaussian that composes the full Wigner function will be translated on average by an amplitude of $\Delta\boldsymbol{\xi}\pp{r}$ depending on its position at each application of the channel. The separate and combined effects of the two noise channels are shown in \cref{fig:noise}.
When $\Delta\boldsymbol{\xi}\pp{r}$ becomes of a length similar to the logical operators of the code, the logical error rate increases. It then becomes important to minimize the radial position of the full finite energy GKP state, which in turn suggests that the position of the envelope $\boldsymbol{\mu}_E$ should be minimized at every step of a computation. Accordingly, we compare different implementations with the metric
\begin{equation}\label{eq:met_mu}
 \text{d}_\mu^2\pp{\boldsymbol{\mu}_E} = \boldsymbol{\mu}_E^T\boldsymbol{\mu}_E.
\end{equation}
More generally, the argument toward reducing the number of excitations in the harmonic oscillator can be used more broadly for most error channels on harmonic oscillators (e.g. bosonic loss, dephasing, spurious Kerr non-linearity), as their effects increase with excitation number. We thus compute the average number of excitations in a GKP state described by generator matrix $S$ and envelope $\pp{\Sigma_E,\boldsymbol{\mu}_E} = \pp{M_E\Sigma_0 M_E^T,\boldsymbol{\mu}_E}$, where $\Sigma_0$ contains the symplectic eigenvalues on the diagonal after a Williamson decomposition \cite{houde_2024}. This result takes the form of a ponderated sum of the excitation contribution over all the Gaussian peaks in phase-space:
\begin{equation}\label{eq:pond_exc_num}
 \left<\hat{\boldsymbol{\xi}}^T\hat{\boldsymbol{\xi}}\right> =  \sum_{\boldsymbol{\chi}_E\in\Lambda_E} f\pp{\boldsymbol{\chi}_E} \pp{ \boldsymbol{\chi}_E^T\boldsymbol{\chi}_E + \text{Tr}\bb{M_E \tanh\pp{\Sigma_0^{-1}}M_E^T} },
\end{equation}
where $f\pp{\boldsymbol{\chi}_E}$ is the ponderation coefficient (see \cref{App_Wig_number} for the full derivation). $\Lambda_E$ is the set of peak positions altered by the envelope. For each $\boldsymbol{\lambda} \,\in\, \Lambda$, we have an associated $\boldsymbol{\chi}_E\,\in\,\Lambda_E$ such that 
\begin{equation}
 \boldsymbol{\chi}_E = M_E\sech\pp{\Sigma_0^{-1}}M_E^{-1}\bb{\boldsymbol{\lambda}-\boldsymbol{\mu}_E} + \boldsymbol{\mu}_E.
\end{equation}
The result of \cref{eq:pond_exc_num} reinforces the criterion of \cref{eq:met_mu} because the envelope position $\boldsymbol{\mu}_E$ that minimizes the contribution 
\begin{equation}
  \sum_{\boldsymbol{\chi}_E} f\pp{\boldsymbol{\chi}_E} \boldsymbol{\chi}_E^T\boldsymbol{\chi}_E,
\end{equation}
also minimizes $\text{d}_\mu^2\pp{\boldsymbol{\mu}_E}$. The second contribution to the excitation number is connected to the covariance matrix of the envelope
\begin{equation}\label{eq:sqz_cont}
  \sum_{\boldsymbol{\chi}_E} f\pp{\boldsymbol{\chi}_E} \text{Tr}\bb{M_E \tanh\pp{\Sigma_0^{-1}}M_E^T}.
\end{equation}
The Williamson decomposition splits the covariance matrix $\Sigma_E$ into the product of a symplectic matrix $M_E$ and a positive definite diagonal matrix $\Sigma_0$: $\Sigma_E = M_E \Sigma_0 M_E^T$. The trace contribution in \cref{eq:sqz_cont} is invariant under orthogonal matrix congruence, leaving only the squeezing part of $M_E$ contributing non-trivially to $\left<\hat{\boldsymbol{\xi}}^T\hat{\boldsymbol{\xi}}\right>$. The contribution increases with the amount of squeezing in $M_E$ and is minimized when $M_E\in {\rm O}_{2N}(\mathbb{R})$ \cite{bhatia_2015}. Altough our general goal is to reduce the excitation number, it is not the only goal of this optimisation. As discussed in \cref{subsec:FEGKP}, the optimal eigenvalues of $\Sigma_0$ depend on the quality of the oscillators and the frequency of error correction. Consequently, reducing the excitation contribution of one mode below its optimal value (at the expense of another mode) means that we are not using the full capacity of the encoding and that errors will lead to quick decoherence. In order to consider squeezing and anti-squeezing on the same footing, we introduce a metric on covariance matrices 
\begin{equation}\label{eq:met_sig}
 \text{d}_{\Sigma}^2\pp{\Sigma_E,\Sigma_0} = \text{Tr}\bb{\ln^2\pp{\sqrt{\Sigma_0^{-1}}\,\Sigma_E \,\sqrt{\Sigma_0^{-1}}}},
\end{equation}
with $\Sigma_E$ the covariance matrix of the envelope and $\Sigma_0$ the optimal covariance matrix from the Williamson decomposition \cite{forstner_2003}. If we consider that all modes are similar, the initial parameters of the envelope are the ones of \cref{eq:start_env}, where $\Sigma_0 = \varepsilon^{-1}I$. Under this similarity assumption, minimizing the cost function 
\begin{equation}\label{eq:ex_met}
  \text{Tr}\bb{ M_EM_E^T}\geq 1,
\end{equation}
would minimize the excitation number of \cref{eq:sqz_cont}, while minimizing the cost function
\begin{equation}\label{eq:dis_met}
  \text{d}_{\Sigma}^2\pp{M_E\pp{\varepsilon^{-1}I}M_E^T,\varepsilon^{-1}I} = \text{Tr}\bb{ \ln^2\pp{M_EM_E^T}}\geq 0,
\end{equation}
would minimize the distance between covariance matrices of \cref{eq:met_sig}. Thus, in the case where $\Sigma_0 = \varepsilon^{-1}I$, both metrics from \cref{eq:ex_met} and \cref{eq:dis_met} lead to the same optimization, which only depends on the squeezing contained in $M_E$.

\noindent Therefore, to reduce the impact of noise channels on GKP states, we aim to reduce the amount of displacement and squeezing that are created by the chosen physical implementation. More general schemes could be designed for multimode GKP by playing with the trade-off of sending errors to one mode to gain fidelity in another. Using such a sacrificial mode could be an interesting optimization technique, but it is outside of the scope of this work. Combining both criteria of \cref{eq:met_mu,eq:met_sig}, we can now compare different physical implementations of a logical Clifford gate, and we are ready to give an explicit description of our compiler algorithm.

\section{\label{sec:gsc} Gaussian stabilizer compiler}

\subsection{Gaussian stabilizer compiler}
As described in the last section, a compiler algorithm selects an implementation given a logical circuit. When considering a physical representative $U_0$ for a logical operation on a code (we choose it closest to the identity for simplicity), $U_0\mathcal{S}$ is the group of all implementations of the same logical operation: $U_0S\ket{\overline{\Psi}} = U_0\ket{\overline{\Psi}} \,\forall \, S\in \mathcal{S}$, where $\mathcal{S}$ is the stabilizer group that can be implemented. Here, we choose $\mathcal S$ to be the gaussian stabilizer group. Any operation logically equivalent to $U_0$ can be written as
\begin{equation}\label{eq:imp_cliff}
 \text{Imp}\pp{\overline{U}} = U_0 \, g_n, 
\end{equation}
where $g_n$ correspond to the application of a sequence of at most $n$ generators of $\mathcal{S}$ (or their inverse). We can look at $g_n$ in \cref{eq:imp_cliff} as a walk of length at most $n$ on the Cayley graph of the Gaussian stabilizer group $ \Gamma\pp{\mathcal{S}_G}$, with generating set as in \cref{eq:gsg}.
A compilation strategy is then translated to a choice of walk on $\Gamma\pp{\mathcal{S}_G}$. A simple compiling strategy is to choose a fixed implementation for a given operation, which we call below the constant compiler. Another compiling strategy, which we refer to as the random walk compiler, is to choose an implementation at random among the closest ones to the identity, by which we mean an implementation of the form of \cref{eq:imp_cliff} with $n=1$. This last strategy induces a random walk on the Cayley graph $\Gamma\pp{\mathcal{S}_G}$ interleaved with the basic physical representatives $\left\{ U_0 \right\}$. As discussed in \cite{conrad_2026}, such a random walk induce a Gaussian stabilizer group twirling, which will result in an approximate logical depolarizing channel. Both the constant and random walk compilers are suboptimal as the amount of squeezing and total displacement increases with the number of operations, as seen in \cref{fig:LRB_curves}. This remains true if we modify the random walk compiler to only sample from a smaller set of generators.

To achieve an optimized circuit implementation, the metrics in \cref{eq:met_mu,eq:met_sig} should be minimized, among all possible walks on $\Gamma\pp{\mathcal{S}_G}$. However, because the Gaussian stabilizer is an infinite group, the metrics cannot be computed for all equivalent operations, and a cutoff needs to be defined. Because operations in $\mathcal{S}_G$ that are far from the identity generate more squeezing and displacement, they are less likely to be useful in the optimization. Therefore, we only consider operations made of short sequences of generators of the stabilizer group. With $\mathcal{S}_G$ finitely generated, we can compute both metrics in \cref{eq:met_mu,eq:met_sig} of all different walks on $\Gamma$ up to a finite length $n$ and define the optimal encoded Clifford implementation as the path that minimizes the metrics at each step. This process is described in \cref{alg:gate_opt}. The algorithm takes as input the target operation $U_0$ and the parameters of the initial envelope ($\Sigma_i$ and $\boldsymbol{\mu}_i$). After comparing the different paths $g_n$ on $\Gamma$ of length $n$, it returns the optimal operation $U_{\text{opt}}$ and the final parameters of the envelope ($\Sigma_{f,\text{opt}}$ and $\boldsymbol{\mu}_{f,\text{opt}}$). We derived an example of the application of \cref{alg:gate_opt} in \cref{app:algo1_ex}. Below, we set the maximum length $n$ to $2$. In general, pass a certain threshold, increasing the walk length $n$ will not necessarily increase the performance of the compiler. A lot of the times, only short walks on $\Gamma$ are used and increasing $n$ will not give better implementations of a gate.
\begin{algorithm}
    \caption{Gaussian stabilizer compiler}\label{alg:gate_opt}
    \begin{algorithmic}
        \REQUIRE $U_0=\hat{T}\pp{\boldsymbol{\lambda}_0}U_{M_0}$, $\Sigma_i$, $\boldsymbol{\mu}_i$ 
        \STATE $\Sigma_{f,\text{opt}} \leftarrow M_0\Sigma_{i}M_0^T$
        \STATE $\boldsymbol{\mu}_{f,\text{opt}} \leftarrow M_0\boldsymbol{\mu}_i + \boldsymbol{\lambda}_0$
        \STATE $g_\text{opt} \leftarrow I$
        \FORALL{ $g_n\,\in\, \left\{g_n\right\} $}
          \STATE Compute its symplectic representation $U_0\, g_n = \hat{T}\pp{\boldsymbol{\lambda}}U_{M}$
            \IF{$d_\Sigma^2\pp{M\Sigma_i M^T, \Sigma_0} <d_\Sigma^2\pp{\Sigma_{f,\text{opt}}, \Sigma_0} $}
            \STATE $\Sigma_{f,\text{opt}} \leftarrow M\Sigma_{i}M^T$
            \STATE $\boldsymbol{\mu}_{f,\text{opt}} \leftarrow M\boldsymbol{\mu}_i + \boldsymbol{\lambda}$
            \STATE $g_\text{opt} \leftarrow g_n$
            \ELSIF{$d_\Sigma^2\pp{M\Sigma_i M^T, \Sigma_0} =d_\Sigma^2\pp{\Sigma_{f,\text{opt}}, \Sigma_0} $ \AND  $d_\mu^2\pp{M\boldsymbol{\mu}_i + \boldsymbol{\lambda}} < d_\mu^2\pp{\boldsymbol{\mu}_{f,opt}}$}
            \STATE $\Sigma_{f,\text{opt}} \leftarrow M\Sigma_{i}M^T$
            \STATE $\boldsymbol{\mu}_{f,\text{opt}} \leftarrow M\boldsymbol{\mu}_i + \boldsymbol{\lambda}$
            \STATE $g_\text{opt} \leftarrow g_n$
            \ENDIF
        \ENDFOR
      \RETURN $U_0\, g_\text{opt}$, $\Sigma_{f,\text{opt}}$, $\boldsymbol{\mu}_{f,\text{opt}}$
    \end{algorithmic}
\end{algorithm}
In \cref{alg:gate_opt}, we choose to optimize the squeezing parameter first and then the displacement since squeezing is typically a larger limiter of the performance of error correction protocols. 
Below, we compare the performance of the Gaussian stabilizer compiler with the constant and random walk compilers and show that it outperforms both of them.

\subsection{Performance of the GS compiler}\label{Perf_GSC}

\noindent A method to compare different implementations of gates at the encoded level is logical randomized benchmarking (LRB) \cite{combes_2017}. Randomized benchmarking (RB) methods are efficient ways to characterize and compare gate set implementations, in a way that the effects of state preparation and measurement errors can be factored out \cite{emerson_2005,knill_2008}. When applying sequences of gates from a group $G$ to a system, the final state is affected compoundly by the decoherence that happens during each gate. The result of the following measurement is then a function of the error channel and the length of the sequence. When the gate implementations are multiplicity-free representations of the target group $G$ and the error model is Markovian and time-independent, the survival probability is well described by a sum of decaying exponentials in the number of operations \cite{helsen_2022}. However, in the setup of LRB, the condition of multiplicity-free representations is often unfulfilled and non-exponential behaviour can appear \cite{ceasura_2022}. In some cases, the survival probability remains a monotonically non-increasing function of the sequence length and can still be used as a measure of performance, but without links to the average gate fidelity. To obtain a measure of performance of the GKP Gaussian compiler, we perform logical randomized benchmarking by sampling uniformly the encoded Clifford group. As encoded Clifford operations are represented by Gaussian operations that are not multiplicity-free, we do not expect to obtain pure exponential decay (at least not for small sequence length \cite{helsen_2022}). Furthermore, the Gaussian stabilizer compiler presented in \cref{alg:gate_opt} does not respect the Markovian condition as it selects gate implementations based on the previous ones in the circuit. Although Markovian noise is a necessary condition for classic RB methods, correlations between the gates, and so between noise applications, are exactly how we gain in performance against decoherence. By using LRB, we do not aim specifically to obtain an error rate for our gate set, but instead aim to compare different compilation strategies for the encoded circuit. An implementation strategy is better than another if, for all circuit lengths, the probability of getting the correct result is greater. Alternatively, it means that longer circuits can be implemented with the same success probability.
\begin{figure}[t]
    \centering
    \includegraphics[width =\linewidth]{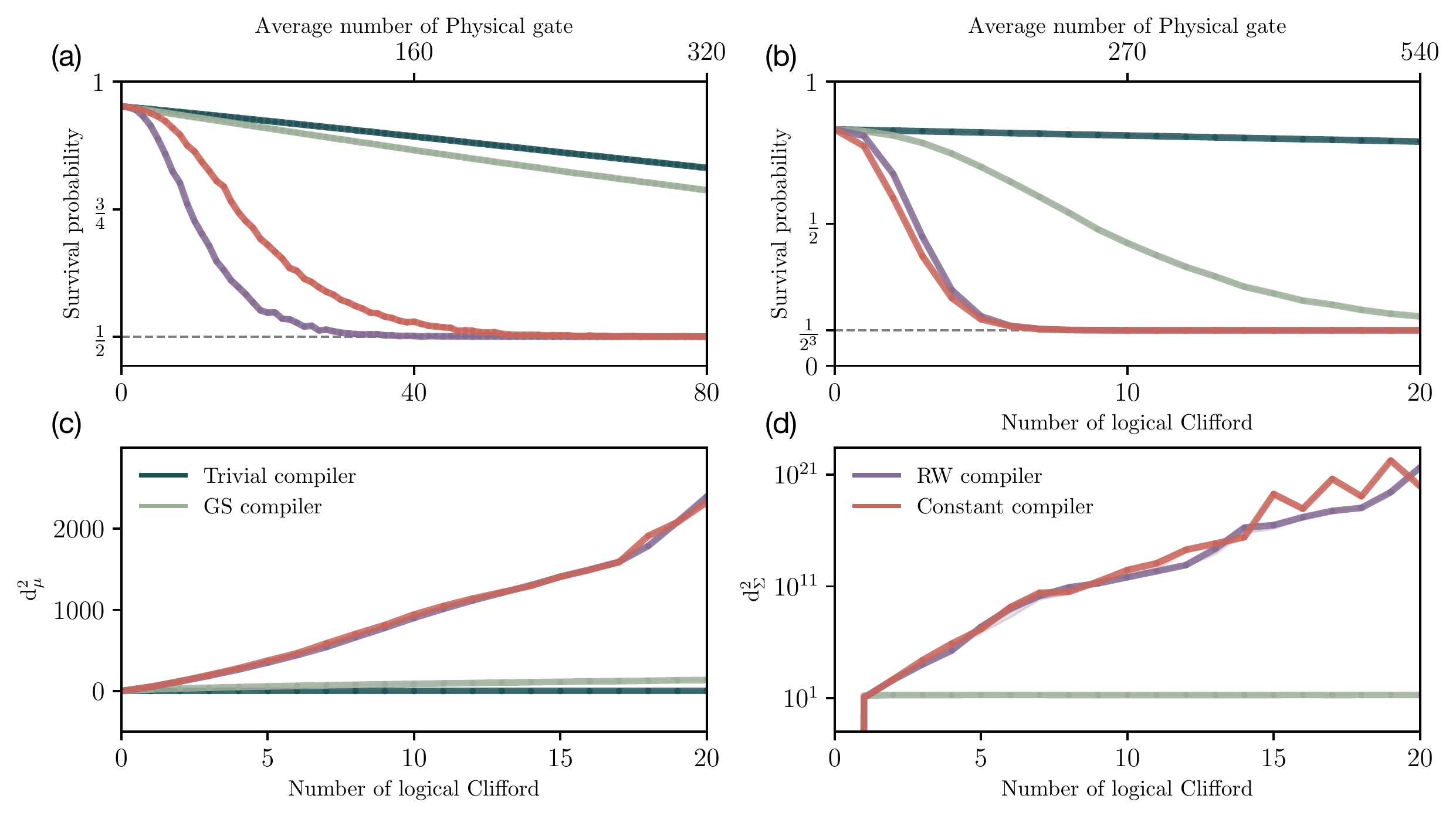}
  \caption{\textbf{Logical randomized benchmarking results.} Survival probability for the constant compiler (red), random walk (purple) and Gaussian stabilizer compiler (light green) on square GKP codes with varying sequence length. We have also included the average survival probability curve when no logical operation is applied (blue-green) as an indicator. The average survival probability curves are shown for (a) 1 and (b) 3 logical qubits with loss strength $\gamma = 0.999$ and $\gamma= 0.9999$, respectively. The dashed line indicates the survival probability for a fully mixed logical state. For the simulations included in the decay curve (b), we compute (c) the maximum position metric of \cref{eq:met_mu} and (d) the maximum squeezing metric of \cref{eq:met_sig} reached as a function of logical circuit length.}
    \label{fig:LRB_curves}
\end{figure}

\noindent With the goal of doing an LRB analysis, we simulate GKP states by following the evolution of their Wigner function in phase-space with the tools developed in Ref.~\cite{bourassa_2021}. In this setup, the Wigner function of a finite energy GKP grid state is described as a sum of Gaussian functions in phase space:
\begin{equation}\label{eq:sum_gaus}
 W_{\text{GKP}}\pp{\xi} = \sum_{m\in \mathcal{M}} c_m G_{\Sigma_m, \boldsymbol{\mu}_m}\pp{\xi},
\end{equation}
with $G_{\Sigma_m, \boldsymbol{\mu}_m}\pp{\xi}$ defined in \cref{eq:gauss_wig} and the set $\mathcal{M}$ indexing the Gaussian functions. The normalization condition reads $\sum_{m\in \mathcal{M}} c_m=1$ for this representation. We have also generalized the results of \cite{bourassa_2021} to obtain a representation of the form of \cref{eq:sum_gaus} for any multimode GKP grid state with potentially squeezed and displaced envelope in \cref{App_Wig_proj,App_Wig_FE_states}. For simplicity, we consider that the system is initialized in a way that all modes are decoupled from each other $\Sigma_E=\varepsilon^{-1} \oplus \varepsilon^{-1}$ and $\boldsymbol{\mu}_E=\boldsymbol{0}$ as in \cref{eq:init_env}. Those considerations lead to an initial state with with parameters
\begin{equation}\label{eq:wig_param}
  \Sigma_m = \frac{1}{4\pi}\coth\pp{\Sigma_E^{-1}}, \qquad \boldsymbol{\mu}_m = \text{sech}\pp{\Sigma_E^{-1}} \boldsymbol{\mu}_{S,m} ,
\end{equation}
where $\boldsymbol{\mu}_{S,m}$ is the position of each Dirac delta function in the infinite energy description of the state with stabilizer generator matrix $S$ derived in \cref{App_Wig_proj}. This description of the state as a sum of Gaussian functions in phase space has many advantages. The most useful being that we can follow its evolution under all Gaussian operations and Gaussian noise channels by updating the parameters in \cref{eq:wig_param}. Under any Gaussian operation, the parameters of the envelope are updated like in \cref{eq:Conj_trans}. This rule can also be used for the dephasing channel of \cref{eq:Depha} if we use a Monte Carlo method to model this channel as a random rotation of each mode. For bosonic loss, the parameters are updated with
\begin{equation}
  \Sigma_m \to \gamma\Sigma_m + \frac{1-\gamma}{4\pi} I,  \qquad \boldsymbol{\mu}_m \to  \sqrt{\gamma} \boldsymbol{\mu}_m,
\end{equation}
when the loss parameter is $\gamma$. In summary, to apply a given Gaussian circuit, we find the representation of the circuit in terms of symplectic operation and apply those symplectic matrices directly onto the saved covariance matrices and peak positions.\\

\noindent At the end of the computation, we wish to apply a decoding gadget to know if a logical error occurs. To achieve this decoding, we compute the characteristic function of a given Wigner function using the Fourier transform. This can be done analytically for any Wigner functions of the form of \cref{eq:sum_gaus} due to the linearity of the transform. The result is again a sum of Gaussian functions that can be evaluated at any point. To complete the connection with GKP codes, we present another form of the characteristic function
\begin{gather}
  \chi\pp{\mathbf{v}} = \text{Tr}\bb{\hat{T}\pp{\mathbf{v}}\rho} = \left< \hat{T}\pp{\mathbf{v}}\right>.
\end{gather}
This form is particularly useful for us because we can obtain the average value of a displacement operators, which include stabilizers of ideal GKP codes. Because we have access to the whole characteristic function analytically, we can efficiently compute an approximation of the projector onto the $+1$ eigenvalue subspace of the logical $\overline{Z}$ operator
\begin{equation}\label{eq:avg_proj}
  \left<\Pi_{\overline{Z}=+1}\right> = \sum_{\mathbf{s}\in\Lambda\cup \Lambda_z} \left<\hat{T}\pp{\mathbf{s}} \right> = \sum_{\mathbf{s}\in\Lambda\cup \Lambda_z} \chi\pp{\mathbf{s}},
\end{equation}
by truncating the sum. The value of \cref{eq:avg_proj} is then interpreted as the probability of having the correct result after the computation (based on an ideal characteristic function decoder), which we also defined as a survival probability. The definition of the projector can be generalized for the operator $\overline{Z}^{\otimes N}$ when multiple qubits are used, again described by a sum of the average value of displacement operators. Hence, we can perform GKP state preparation, state evolution under logical Clifford circuits and survival probability measurements efficiently.\\

\noindent To analyze the advantages of using the Gaussian stabilizer compiler over other compiler algorithms, we performed LRB simulations for 1,2 and 3 logical qubits encoded in a square GKP code. The LRB simulations consist of generating a large number of logical Clifford circuits varying in length and simulating the implementation of those circuits, obtained from a given compiler algorithm. By averaging the survival probability obtained over all random circuits of a given length, we obtain the average performance of a compiler at this length. The results for 1 and 3 qubits are showcased in \cref{fig:LRB_curves} in panels a and b, respectively. As expected, the decay curves do not exhibit the regular exponential behaviour for short sequences, but rather a Gaussian behaviour. The Gaussian behaviour can be explained by the shape of the characteristic function of the finite-energy GKP states. This function, like its Wigner function, is composed of a sum of many Gaussian features. The effect of the noise channel is to shift the peaks of the Gaussian away from the measured points, leading to a reduction in the survival probability. For short sequences of gates, the resulting shape of the decay then corresponds to the shape of the characteristic function near the measured point. For longer logical circuits, logical Pauli errors become more important, and we recover the exponential decay behaviour. As discussed at the beginning of the section, even if the decay does not follow an exponential behaviour, we still obtain a lifetime ordering of the different compilation strategies. 

\Cref{fig:LRB_curves} demonstrates that the constant and random walk compilers have similar performance, while the Gaussian stabilizer compiler performs better. In this situation, we expected the constant and random compiler to perform similarly. Since the logical Hadamard gate is implemented via a $\pi/2$ phase-space rotation, it leads to an inversion of coordinates in phase-space, which means that the randomness from the circuit sampling step can also be understood as a random walk, leading to similar performance as the random walk compiler. We observe that in all cases, the Gaussian stabilizer compiler performs better than the other two compilers. We also observe that, as expected, both the displacement and squeezing metrics remain low during computations with the GS compiler.
\begin{figure}[t]
    \centering
    \includegraphics[width =\linewidth]{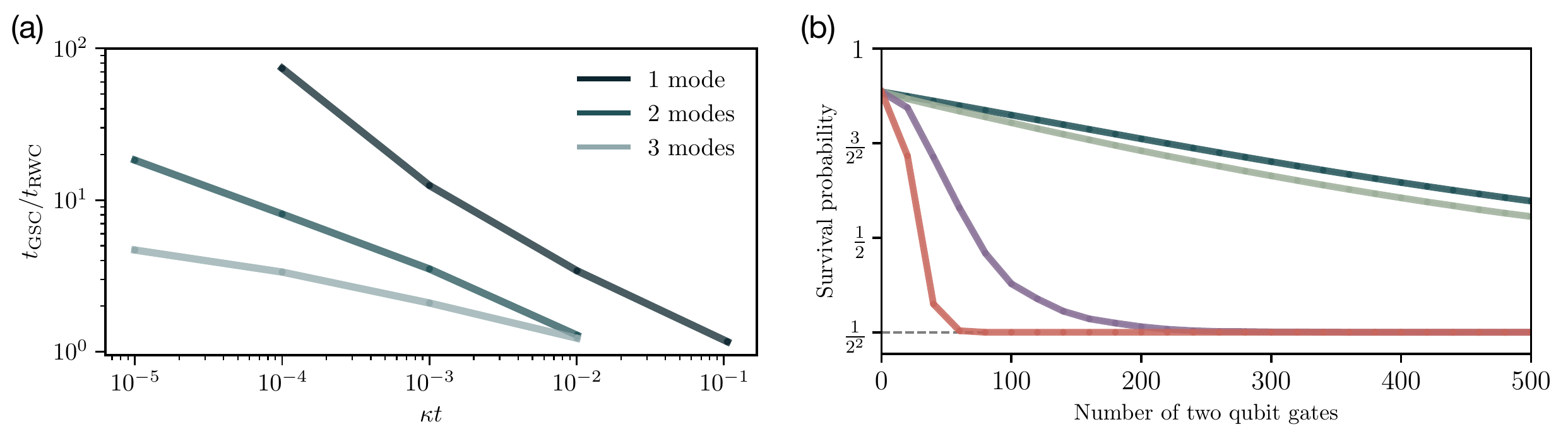}
  \caption{\textbf{Performance of the GS compiler with varying randomized benchmarking conditions.} (a) Lifetime amelioration factor between the Gaussian stabilizer compiler and random walk compiler as a function of loss rate ($\kappa t$). (b) Results of two-qubit gates randomized benchmarking with the group $\left<\overline{CX},\overline{CZ}\right>$ and loss parameter $\gamma=0.999$. The figure shows the survival probability for the constant compiler (red), random walk (purple) and Gaussian stabilizer compiler (light green) on two logical square GKP codes with varying sequence length.}
    \label{fig:Am_noise}
\end{figure} 
To extract a computation lifetime, we fit a Gaussian function on the survival probability $P(x)$
\begin{equation}
  P\pp{x} = A\mathrm{e}^{-\pp{ax^2+bx}}+B,
\end{equation}
where $x$ is the number of logical gates applied. For all decay curves, we define the computation lifetime $t$ as the number of logical operations that we need to apply to reduce the survival probability by a factor of $1/\mathrm{e}$: 
\begin{equation}
  P\pp{t} = A\mathrm{e}^{-1}+B.
\end{equation}
This definition of the lifetime allows us to compare both Gaussian and exponential fits on an equal footing. The amelioration factor, defined as the ratio of the GS compiler lifetime over the random walk compiler lifetime, is dependent on bosonic loss strength. As shown in \cref{fig:Am_noise} (a), we obtain that, at low loss rate, the GS compiler performs even better. When the loss rate is high, no compiler can help preserve the logical information and the computation fails with high probability in all cases. Another result from this simulation is that the amelioration factor seems to decrease with the number of modes. We explain this behaviour by the fact that, as we increase the number of modes and therefore the number of logical Clifford gates that we cannot simplify, the state spends a lot of time in moderatly squeezed state before we can actually apply any simplification. And so, as we increase the number of modes, we need to reduce the noise strength to achieve the same performance.

Finally, we performed similar LRB simulations, but now replacing the full Clifford group with two-qubit gates only. This scenario is relevant in a situation where a quantum compution is performed with magic state injection and Clifford gates. In this case, all single-qubit Clifford gates can be performed in a Pauli and Clifford frame, leaving only two-qubit gates to be implemented physically. As a result, only squeezing would play a role here, since all displacement operations would be removed. The result of a logical randomized benchmarking simulation with only two-qubit gates on two logical qubits is included in \cref{fig:Am_noise} (b). We find that the Gaussian stabilizer compiler still performs well in this setup compared to other compiling strategies.

As a final note, we remark that interleaving quantum error correction gadgets in GKP logical circuits will have an effect on the envelope and might affect the results presented here. For example, error correction via state teleportation resets the envelope of the state, the sBs protocol stabilizes a centered envelope, and displacement twirling of the envelope reduce the logical space leakage \cite{conrad_2026,jafarzadeh_2025}. However, we believe that our Gaussian stabilizer compiler will remain useful with hybrid approaches depending on the actual implementation of error correction. We leave the analysis of combining quantum error correction schemes with the GS compiler to future work.

\section{\label{sec:C}Conclusion}
In this work, we have investigated how to leverage more symmetries of GKP codes to reduce the impact of noise at the logical level. We first presented how we can obtain more symmetries of a code by relaxing the restriction to abelian stabilizer groups. This new perspective enables us to find all symmetry operations related to an implementation of a logical operator. Restricting this set to Gaussian operations, we showed that it is possible to describe the Gaussian symmetry group with only a few generators for any lattice describing a GKP code. This new set of non-commuting stabilizers can be used similarly to the commuting ones, such as for state stabilization, quantum error correction and decoding. In future work, we aim to characterize more thoroughly this connection between non-abelian stabilizer groups and potential applications in quantum error correction.

Exploring the passage from infinite to finite energy GKP state, we found that infinite energy stabilizers become logical stabilizers, but not physical stabilizers. As a first application of GKP non-abelian stabilizer groups, we presented a compiler algorithm that uses those logical symmetries on the finite-energy version of GKP codes to reduce the impact of loss and dephasing. This compiler works by selecting an implementation of a logical operation that reduces as much as possible the amount of total displacement and squeezing. Doing so reduces the mean excitation number in the oscillator and, by consequence, reduces the impact of loss and dephasing. We conjecture that it should also increase the protection against other noise channels, such as Kerr non-linearity. We leave that analysis for future work. It would also be interesting to explore how non-Gaussian stabilizer operations can fit within this formalism and increase their fidelity.

We showed that we obtain an amelioration in computation lifetime while using the Gaussian stabilizer compiler, and that this amelioration increases with the reduction of loss strength. While the GS compiler only uses the GKP symmetries to recenter and unsqueezed the intermediate states of a computation, it seems probable that dynamical decoupling techniques at the compilation level could help symmetrize the effects of noise and further reduce its impact.

It is conceivable that GKP codes will be used as a base layer for a fault-tolerant quantum computer, with a top-layer qubit code \cite{noh_2022}. To reach this goal, it is primordial to have good native operations on the base layer combined with a performant compilation strategy. In this paper, we have introduced one such strategy. 

\begin{acknowledgments}
This research was undertaken thanks in part to
funding from the Canada First Research Excellence
Fund. This research was also funded by the
National Science and Engineering Council with a
Canada Graduate Research Scholarship –
Doctoral, by the Army Research Office under grant
\#W911NF2310045 and by a Université de
Sherbrooke Research Excellence Scholarships.
\end{acknowledgments}

\section*{Code Availability}
The source code for the Gaussian stabilizer compiler and the simulations of logical randomized benchmarking is available at \url{https://github.com/Jopelxl/Gaussian-Stabilizer-Compiler.git}.

\bibliographystyle{quantum}

\begin{thebibliography}{10}

\bibitem{sivak_2023}
V.~V. Sivak, A.~Eickbusch, B.~Royer, S.~Singh, I.~Tsioutsios, S.~Ganjam, A.~Miano, B.~L. Brock, A.~Z. Ding, L.~Frunzio, S.~M. Girvin, R.~J. Schoelkopf, and M.~H. Devoret.
\newblock ``Real-time quantum error correction beyond break-even''.
\newblock \href{https://dx.doi.org/10.1038/s41586-023-05782-6}{Nature {\bf 616}, 50--55}~(2023).
\newblock  \href{http://arxiv.org/abs/2211.09116}{arXiv:2211.09116}.

\bibitem{lachance-quirion_2024}
Dany {Lachance-Quirion}, Marc-Antoine Lemonde, Jean~Olivier Simoneau, Lucas {St-Jean}, Pascal Lemieux, Sara Turcotte, Wyatt Wright, Am{\'e}lie Lacroix, Jo{\"e}lle {Fr{\'e}chette-Viens}, Ross Shillito, Florian Hopfmueller, Maxime Tremblay, Nicholas~E. Frattini, Julien Camirand~Lemyre, and Philippe {St-Jean}.
\newblock ``Autonomous {{Quantum Error Correction}} of {{Gottesman-Kitaev-Preskill States}}''.
\newblock \href{https://dx.doi.org/10.1103/PhysRevLett.132.150607}{Physical Review Letters {\bf 132}, 150607}~(2024).

\bibitem{brock_2025}
Benjamin~L. Brock, Shraddha Singh, Alec Eickbusch, Volodymyr~V. Sivak, Andy~Z. Ding, Luigi Frunzio, Steven~M. Girvin, and Michel~H. Devoret.
\newblock ``Quantum error correction of qudits beyond break-even''.
\newblock \href{https://dx.doi.org/10.1038/s41586-025-08899-y}{Nature {\bf 641}, 612--618}~(2025).

\bibitem{ofek_2016}
Nissim Ofek, Andrei Petrenko, Reinier Heeres, Philip Reinhold, Zaki Leghtas, Brian Vlastakis, Yehan Liu, Luigi Frunzio, S.~M. Girvin, L.~Jiang, Mazyar Mirrahimi, M.~H. Devoret, and R.~J. Schoelkopf.
\newblock ``Extending the lifetime of a quantum bit with error correction in superconducting circuits''.
\newblock \href{https://dx.doi.org/10.1038/nature18949}{Nature {\bf 536}, 441--445}~(2016).

\bibitem{konno_2024}
Shunya Konno, Warit Asavanant, Fumiya Hanamura, Hironari Nagayoshi, Kosuke Fukui, Atsushi Sakaguchi, Ryuhoh Ide, Fumihiro China, Masahiro Yabuno, Shigehito Miki, Hirotaka Terai, Kan Takase, Mamoru Endo, Petr Marek, Radim Filip, Peter {van Loock}, and Akira Furusawa.
\newblock ``Logical states for fault-tolerant quantum computation with propagating light''.
\newblock \href{https://dx.doi.org/10.1126/science.adk7560}{Science {\bf 383}, 289--293}~(2024).

\bibitem{larsen_2025}
M.~V. Larsen, J.~E. Bourassa, S.~Kocsis, J.~F. Tasker, R.~S. Chadwick, C.~{Gonz{\'a}lez-Arciniegas}, J.~Hastrup, C.~E. {Lopetegui-Gonz{\'a}lez}, F.~M. Miatto, A.~Motamedi, R.~Noro, G.~Roeland, R.~Baby, H.~Chen, P.~Contu, I.~Di~Luch, C.~Drago, M.~Giesbrecht, T.~Grainge, I.~Krasnokutska, M.~Menotti, B.~Morrison, C.~Puviraj, K.~Rezaei~Shad, B.~Hussain, J.~McMahon, J.~E. Ortmann, M.~J. Collins, C.~Ma, D.~S. Phillips, M.~Seymour, Q.~Y. Tang, B.~Yang, Z.~Vernon, R.~N. Alexander, and D.~H. Mahler.
\newblock ``Integrated photonic source of {{Gottesman}}--{{Kitaev}}--{{Preskill}} qubits''.
\newblock \href{https://dx.doi.org/10.1038/s41586-025-09044-5}{Nature {\bf 642}, 587--591}~(2025).

\bibitem{deneeve_2022}
Brennan {de Neeve}, Thanh-Long Nguyen, Tanja Behrle, and Jonathan~P. Home.
\newblock ``Error correction of a logical grid state qubit by dissipative pumping''.
\newblock \href{https://dx.doi.org/10.1038/s41567-021-01487-7}{Nature Physics {\bf 18}, 296--300}~(2022).

\bibitem{matsos_2024a}
V.~G. Matsos, C.~H. Valahu, T.~Navickas, A.~D. Rao, M.~J. Millican, X.~C. Kolesnikow, M.~J. Biercuk, and T.~R. Tan.
\newblock ``Robust and {{Deterministic Preparation}} of {{Bosonic Logical States}} in a {{Trapped Ion}}''.
\newblock \href{https://dx.doi.org/10.1103/PhysRevLett.133.050602}{Physical Review Letters {\bf 133}, 050602}~(2024).

\bibitem{matsos_2025}
V.~G. Matsos, C.~H. Valahu, M.~J. Millican, T.~Navickas, X.~C. Kolesnikow, M.~J. Biercuk, and T.~R. Tan.
\newblock ``Universal quantum gate set for {{Gottesman}}--{{Kitaev}}--{{Preskill}} logical qubits''.
\newblock \href{https://dx.doi.org/10.1038/s41567-025-03002-8}{Nature PhysicsPages 1--6}~(2025).

\bibitem{gottesman_2001}
Daniel Gottesman, Alexei Kitaev, and John Preskill.
\newblock ``Encoding a qubit in an oscillator''.
\newblock \href{https://dx.doi.org/10.1103/PhysRevA.64.012310}{Physical Review A {\bf 64}, 012310}~(2001).
\newblock  \href{http://arxiv.org/abs/quant-ph/0008040}{arXiv:quant-ph/0008040}.

\bibitem{noh_2019}
Kyungjoo Noh, Victor~V. Albert, and Liang Jiang.
\newblock ``Quantum {{Capacity Bounds}} of {{Gaussian Thermal Loss Channels}} and {{Achievable Rates With Gottesman-Kitaev-Preskill Codes}}''.
\newblock \href{https://dx.doi.org/10.1109/TIT.2018.2873764}{IEEE Transactions on Information Theory {\bf 65}, 2563--2582}~(2019).

\bibitem{leviant_2022}
Peter Leviant, Qian Xu, Liang Jiang, and Serge Rosenblum.
\newblock ``Quantum capacity and codes for the bosonic loss-dephasing channel''.
\newblock \href{https://dx.doi.org/10.22331/q-2022-09-29-821}{Quantum {\bf 6}, 821}~(2022).
\newblock  \href{http://arxiv.org/abs/2205.00341}{arXiv:2205.00341}.

\bibitem{bravyi_2005}
Sergey Bravyi and Alexei Kitaev.
\newblock ``Universal quantum computation with ideal {{Clifford}} gates and noisy ancillas''.
\newblock \href{https://dx.doi.org/10.1103/PhysRevA.71.022316}{Physical Review A {\bf 71}, 022316}~(2005).

\bibitem{royer_2020}
Baptiste Royer, Shraddha Singh, and S.~M. Girvin.
\newblock ``Stabilization of {{Finite-Energy Gottesman-Kitaev-Preskill States}}''.
\newblock \href{https://dx.doi.org/10.1103/PhysRevLett.125.260509}{Physical Review Letters {\bf 125}, 260509}~(2020).
\newblock  \href{http://arxiv.org/abs/2009.07941}{arXiv:2009.07941}.

\bibitem{tzitrin_2020}
Ilan Tzitrin, J.~Eli Bourassa, Nicolas~C. Menicucci, and Krishna~Kumar Sabapathy.
\newblock ``Progress towards practical qubit computation using approximate {{Gottesman-Kitaev-Preskill}} codes''.
\newblock \href{https://dx.doi.org/10.1103/PhysRevA.101.032315}{Physical Review A {\bf 101}, 032315}~(2020).

\bibitem{harrington_2001}
Jim Harrington and John Preskill.
\newblock ``Achievable rates for the {{Gaussian}} quantum channel''.
\newblock \href{https://dx.doi.org/10.1103/PhysRevA.64.062301}{Physical Review A {\bf 64}, 062301}~(2001).

\bibitem{conrad_2022}
Jonathan Conrad, Jens Eisert, and Francesco Arzani.
\newblock ``Gottesman-{{Kitaev-Preskill}} codes: {{A}} lattice perspective''.
\newblock \href{https://dx.doi.org/10.22331/q-2022-02-10-648}{Quantum {\bf 6}, 648}~(2022).
\newblock  \href{http://arxiv.org/abs/2109.14645}{arXiv:2109.14645}.

\bibitem{royer_2022}
Baptiste Royer, Shraddha Singh, and Steven~M. Girvin.
\newblock ``Encoding qubits in multimode grid states''.
\newblock \href{https://dx.doi.org/10.1103/PRXQuantum.3.010335}{PRX Quantum {\bf 3}, 010335}~(2022).
\newblock  \href{http://arxiv.org/abs/2201.12337}{arXiv:2201.12337}.

\bibitem{gottesman_1997}
Daniel Gottesman.
\newblock ``Stabilizer {{Codes}} and {{Quantum Error Correction}}, {{Caltech Ph}}.{{D}}. thesis''~(1997).
\newblock  \href{http://arxiv.org/abs/quant-ph/9705052}{arXiv:quant-ph/9705052}.

\bibitem{ni_2015}
Xiaotong Ni, Oliver Buerschaper, and Maarten {Van den Nest}.
\newblock ``A non-commuting stabilizer formalism''.
\newblock \href{https://dx.doi.org/10.1063/1.4920923}{Journal of Mathematical Physics {\bf 56}, 052201}~(2015).

\bibitem{webster_2022}
Mark~A. Webster, Benjamin~J. Brown, and Stephen~D. Bartlett.
\newblock ``The {{XP Stabiliser Formalism}}: A {{Generalisation}} of the {{Pauli Stabiliser Formalism}} with {{Arbitrary Phases}}''.
\newblock \href{https://dx.doi.org/10.22331/q-2022-09-22-815}{Quantum {\bf 6}, 815}~(2022).

\bibitem{boudreault_2025}
J{\'e}r{\'e}mie Boudreault, Ross Shillito, Jean-Baptiste Bertrand, and Baptiste Royer.
\newblock ``Using a {{Kerr}} interaction for {{GKP}} magic state preparation''~(2025).
\newblock  \href{http://arxiv.org/abs/2507.09684}{arXiv:2507.09684}.

\bibitem{rengaswamy_2020}
Narayanan Rengaswamy, Robert Calderbank, Swanand Kadhe, and Henry~D. Pfister.
\newblock ``Logical {{Clifford Synthesis}} for {{Stabilizer Codes}}''.
\newblock \href{https://dx.doi.org/10.1109/TQE.2020.3023419}{IEEE Transactions on Quantum Engineering {\bf 1}, 1--17}~(2020).

\bibitem{kuehnke_2025}
Eric~J. Kuehnke, Kyano Levi, Joschka Roffe, Jens Eisert, and Daniel Miller.
\newblock ``Hardware-tailored logical {{Clifford}} circuits for stabilizer codes''~(2025).
\newblock  \href{http://arxiv.org/abs/2505.20261}{arXiv:2505.20261}.

\bibitem{conrad_2024}
Jonathan Conrad, Ansgar~G. Burchards, and Steven~T. Flammia.
\newblock ``Lattices, {{Gates}}, and {{Curves}}: {{GKP}} codes as a {{Rosetta}} stone''~(2024).
\newblock  \href{http://arxiv.org/abs/2407.03270}{arXiv:2407.03270}.

\bibitem{lin_2023}
Mao Lin, Christopher Chamberland, and Kyungjoo Noh.
\newblock ``Closest {{Lattice Point Decoding}} for {{Multimode Gottesman-Kitaev-Preskill Codes}}''.
\newblock \href{https://dx.doi.org/10.1103/PRXQuantum.4.040334}{PRX Quantum {\bf 4}, 040334}~(2023).

\bibitem{weedbrook_2012}
Christian Weedbrook, Stefano Pirandola, Ra{\'u}l {Garc{\'i}a-Patr{\'o}n}, Nicolas~J. Cerf, Timothy~C. Ralph, Jeffrey~H. Shapiro, and Seth Lloyd.
\newblock ``Gaussian quantum information''.
\newblock \href{https://dx.doi.org/10.1103/RevModPhys.84.621}{Reviews of Modern Physics {\bf 84}, 621--669}~(2012).

\bibitem{houde_2024}
Martin Houde, Will McCutcheon, and Nicol{\'a}s Quesada.
\newblock ``Matrix decompositions in quantum optics: {{Takagi}}/{{Autonne}}, {{Bloch}}--{{Messiah}}/{{Euler}}, {{Iwasawa}}, and {{Williamson}}''.
\newblock \href{https://dx.doi.org/10.1139/cjp-2024-0070}{Canadian Journal of Physics {\bf 102}, 497--507}~(2024).

\bibitem{bhatia_2015}
Rajendra Bhatia and Tanvi Jain.
\newblock ``On symplectic eigenvalues of positive definite matrices''.
\newblock \href{https://dx.doi.org/10.1063/1.4935852}{Journal of Mathematical Physics {\bf 56}, 112201}~(2015).
\newblock  \href{http://arxiv.org/abs/1803.04647}{arXiv:1803.04647}.

\bibitem{forstner_2003}
Wolfgang F{\"o}rstner and Boudewijn Moonen.
\newblock ``A {{Metric}} for {{Covariance Matrices}}''.
\newblock In Erik~W. Grafarend, Friedrich~W. Krumm, and Volker~S. Schwarze, editors, Geodesy-{{The Challenge}} of the 3rd {{Millennium}}.
\newblock \href{https://dx.doi.org/10.1007/978-3-662-05296-9_31}{Pages 299--309}.
\newblock Springer Berlin Heidelberg, Berlin, Heidelberg~(2003).

\bibitem{conrad_2026}
Jonathan Conrad, Jens Eisert, and Steven~T. Flammia.
\newblock ``Chasing shadows with {{Gottesman-Kitaev-Preskill}} codes''.
\newblock \href{https://dx.doi.org/10.22331/q-2026-01-19-1973}{Quantum {\bf 10}, 1973}~(2026).

\bibitem{combes_2017}
Joshua Combes, Christopher Granade, Christopher Ferrie, and Steven~T. Flammia.
\newblock ``Logical {{Randomized Benchmarking}}''~(2017).
\newblock  \href{http://arxiv.org/abs/1702.03688}{arXiv:1702.03688}.

\bibitem{emerson_2005}
Joseph Emerson, Robert Alicki, and Karol {\.Z}yczkowski.
\newblock ``Scalable noise estimation with random unitary operators''.
\newblock \href{https://dx.doi.org/10.1088/1464-4266/7/10/021}{Journal of Optics B: Quantum and Semiclassical Optics {\bf 7}, S347}~(2005).

\bibitem{knill_2008}
E.~Knill, D.~Leibfried, R.~Reichle, J.~Britton, R.~B. Blakestad, J.~D. Jost, C.~Langer, R.~Ozeri, S.~Seidelin, and D.~J. Wineland.
\newblock ``Randomized benchmarking of quantum gates''.
\newblock \href{https://dx.doi.org/10.1103/PhysRevA.77.012307}{Physical Review A {\bf 77}, 012307}~(2008).

\bibitem{helsen_2022}
J.~Helsen, I.~Roth, E.~Onorati, A.H. Werner, and J.~Eisert.
\newblock ``General {{Framework}} for {{Randomized Benchmarking}}''.
\newblock \href{https://dx.doi.org/10.1103/PRXQuantum.3.020357}{PRX Quantum {\bf 3}, 020357}~(2022).

\bibitem{ceasura_2022}
Athena Ceasura, Pavithran Iyer, Joel~J. Wallman, and Hakop Pashayan.
\newblock ``Non-{{Exponential Behaviour}} in {{Logical Randomized Benchmarking}}''~(2022).
\newblock  \href{http://arxiv.org/abs/2212.05488}{arXiv:2212.05488}.

\bibitem{bourassa_2021}
J.~Eli Bourassa, Nicol{\'a}s Quesada, Ilan Tzitrin, Antal Sz{\'a}va, Theodor Isacsson, Josh Izaac, Krishna~Kumar Sabapathy, Guillaume Dauphinais, and Ish Dhand.
\newblock ``Fast simulation of bosonic qubits via {{Gaussian}} functions in phase space''.
\newblock \href{https://dx.doi.org/10.1103/PRXQuantum.2.040315}{PRX Quantum {\bf 2}, 040315}~(2021).
\newblock  \href{http://arxiv.org/abs/2103.05530}{arXiv:2103.05530}.

\bibitem{jafarzadeh_2025}
Mahnaz Jafarzadeh, Jonathan Conrad, Rafael~N. Alexander, and Ben~Q. Baragiola.
\newblock ``Logical channels in approximate {{Gottesman-Kitaev-Preskill}} error correction''.
\newblock \href{https://dx.doi.org/10.1103/c8hk-v1qf}{Physical Review A {\bf 112}, 062413}~(2025).

\bibitem{noh_2022}
Kyungjoo Noh, Christopher Chamberland, and Fernando~G.S.L. Brand{\~a}o.
\newblock ``Low-{{Overhead Fault-Tolerant Quantum Error Correction}} with the {{Surface-GKP Code}}''.
\newblock \href{https://dx.doi.org/10.1103/PRXQuantum.3.010315}{PRX Quantum {\bf 3}, 010315}~(2022).

\bibitem{mumford_2007a}
David Mumford.
\newblock ``Tata {{Lectures}} on {{Theta I}}''.
\newblock \href{https://dx.doi.org/10.1007/978-0-8176-4577-9}{Modern {{Birkhauser Classics}}}. Birkhauser Boston. Boston, MA~(2007).

\bibitem{baker_1958}
George~A. Baker.
\newblock ``Formulation of {{Quantum Mechanics Based}} on the {{Quasi-Probability Distribution Induced}} on {{Phase Space}}''.
\newblock \href{https://dx.doi.org/10.1103/PhysRev.109.2198}{Physical Review {\bf 109}, 2198--2206}~(1958).

\bibitem{carmichael_1999}
Howard~J. Carmichael.
\newblock ``Statistical {{Methods}} in {{Quantum Optics}} 1''.
\newblock \href{https://dx.doi.org/10.1007/978-3-662-03875-8}{Springer Berlin Heidelberg}. Berlin, Heidelberg~(1999).

\end{thebibliography}

\appendix

\section{Derivation of gaussian stabilizer}\label{App_Symp_demo}
Starting from \cref{iso_eq}, we derive in this section the condition on $M$ such that it defines a symplectic automorphism of the grid. In \cref{GSG}, we derived that the vector $\pp{M-I}\boldsymbol{p}_j$ must be part of the grid $\Lambda$. A general element $\boldsymbol{p}_0$ of $\Lambda$ can always be written as $ \boldsymbol{p}_0 = S^T \mathbf{a} \:\text{for}\: \mathbf{a}\in\mathbb{Z}^{2N}$. At the same time, a vector $\boldsymbol{p}_j$ of the dual grid can always be written as $\boldsymbol{p}_j = \Omega S^{-1}\mathbf{b}\:\text{for}\: \mathbf{b}\in\mathbb{Z}^{2N}$. \Cref{iso_eq} is modified to
\begin{equation}\label{same_vector}
 \pp{M-I}\Omega S^{-1}\mathbf{b} = S^T \mathbf{a}.
\end{equation}
There always exist a matrix $X\in \text{Mat}_{2N}\pp{\mathbb{Z}}$, potentially singular, such that $\mathbf{a} = -X\mathbf{b}$. \Cref{same_vector} then becomes 
\begin{equation}
 \pp{M-I}\Omega S^{-1}\mathbf{b} = -S^T X\mathbf{b} \implies M = S^T XS\Omega+I  .
\end{equation}
But we know that $A = S\Omega S^T$, from which we have the form of \cref{symp_stab_def}
\begin{equation}
 M = S^T\pp{XA+I}S^{-T}.
\end{equation}
To force this matrix to be symplectic, we need that $M\Omega M^T = \Omega$. This condition becomes 
\begin{equation}
 S^T\pp{XA+I}S^{-T}\Omega S^{-1}\pp{A^TX^T+I}S = \Omega.
\end{equation}
Using the relation $A=S\Omega S^T$ and the skew-symmetry of $A$, we arrive at the conclusion of \cref{XA_equation}:
\begin{equation}
 XAX^T - X + X^T =0.
\end{equation} 

\section{Fixing the translation part of Gaussian stabilizers}\label{app:restriction}
In this appendix, we derive the set of $2N$ equations obtained from \cref{eq:rest} that need to be solve to obtain $\boldsymbol{\lambda}^*\in\Lambda^*$ in the transformation $\mathcal{L}_{M,\boldsymbol{\lambda}^*}$. Starting with the restriction from \cref{eq:rest}
\begin{equation}
  \hat{T}\left(\boldsymbol{\lambda}^*\right)U_M \hat{T}\left(\mathbf{p}_j\right)\ket{\psi} = \hat{T}\left(\mathbf{p}_j\right) \hat{T}\left(\boldsymbol{\lambda}^*\right)U_M \ket{\psi}, 
\end{equation}
we can permute the translation by the Pauli representative $\mathbf{p}_j$ with the other two operators on the left side, at the cost of a multiplication by $M$ on the argument
\begin{equation}
  \hat{T}\left(\boldsymbol{\lambda}^*\right)\hat{T}\left(M \mathbf{p}_j\right)U_M \ket{\psi} = \hat{T}\left(\mathbf{p}_j\right) \hat{T}\left(\boldsymbol{\lambda}^*\right)U_M \ket{\psi}, 
\end{equation}
and an additional phase
\begin{equation}
  \exp{-2\pi i \boldsymbol{\lambda}^{*T}\Omega M\mathbf{p}_j}\hat{T}\left(M \mathbf{p}_j\right)\hat{T}\left(\boldsymbol{\lambda}^*\right)U_M \ket{\psi} = \hat{T}\left(\mathbf{p}_j\right) \hat{T}\left(\boldsymbol{\lambda}^*\right)U_M \ket{\psi}.
\end{equation}
Because we are asking that $\mathcal{L}_{M,\boldsymbol{\lambda}^*}$ acts as a stabilizer, we can reduce the last equation to
\begin{equation}
  \exp{-2\pi i \boldsymbol{\lambda}^{*T}\Omega M\mathbf{p}_j}\hat{T}\left(M \mathbf{p}_j\right) \ket{\psi} = \hat{T}\left(\mathbf{p}_j\right)  \ket{\psi}.
\end{equation}
At this point, we can reframe the argument of the translation by $M\mathbf{p}_j$ as a translation by $\bb{M-I}\mathbf{p}_j+\mathbf{p}_j$
\begin{equation}
    \exp{-2\pi i \boldsymbol{\lambda}^{*T}\Omega M\mathbf{p}_j}\hat{T}\left(\bb{M-I}\mathbf{p}_j+\mathbf{p}_j\right) \ket{\psi} = \hat{T}\left(\mathbf{p}_j\right)  \ket{\psi},
\end{equation}
and split it into two translation operators using \cref{eq:trans_relation}
\begin{equation}\label{eq:big}
    \exp{-2\pi i \boldsymbol{\lambda}^{*T}\Omega M\mathbf{p}_j-\pi i \mathbf{p}_j^T\Omega\bb{M-I}\mathbf{p}_j}\hat{T}\left(\mathbf{p}_j\right) \hat{T}\left(\bb{M-I}\mathbf{p}_j\right)\ket{\psi} = \hat{T}\left(\mathbf{p}_j\right)  \ket{\psi}.
\end{equation}
With the definition of the matrix $M$, we know that $\bb{M-I}\mathbf{p}_j\in\Lambda$ (see \cref{iso_eq}) and so, for trivial gauge, we have
\begin{equation}\label{eq:stab_ele}
  \exp{\pi i \mathbf{p}_j^T\bb{M-I}^TS^{-1}A_\text{\lltriangle}S^{-T}\bb{M-I}\mathbf{p}_j}\hat{T}\left(\bb{M-I}\mathbf{p}_j\right) \in \mathcal{S},
\end{equation}
where $A_\text{\lltriangle}$ is the lower triangular part of the symplectic Gram matrix $A=S\Omega S^T$ and $\boldsymbol{\mu}$ is the gauge vector. With the identification of a stabilizer element in \cref{eq:stab_ele}, we can reduce \cref{eq:big} to 
\begin{equation}
  \mathrm{e}^{-i\pi \pp{ 2\boldsymbol{\lambda}^{*T}\Omega M\mathbf{p}_j +  \mathbf{p}_j^T\Omega\bb{M-I}\mathbf{p}_j +  \mathbf{p}_j^T\bb{M-I}^TS^{-1}A_\text{\lltriangle}S^{-T}\bb{M-I}\mathbf{p}_j           }}\hat{T}\left(\mathbf{p}_j\right) \ket{\psi} = \hat{T}\left(\mathbf{p}_j\right)  \ket{\psi}.
\end{equation}
The solutions of this equation are the vectors $\boldsymbol{\lambda}^*$ for which the phase is the identity $\forall j$. What we get is $2N$ equations (one for each generator $\mathbf{p}_j$ of the dual lattice) of the form
\begin{equation}
  2\boldsymbol{\lambda}^{*T}\Omega M\mathbf{p}_j + c_j = 0 \mod 2,
\end{equation}
which can always be solve since the generators $\mathbf{p}_j$ are linearly independant.

\section{Generalized multimode sBs protocol}\label{app_sBs}
In this section, we derive a generalized sBs protocol to stabilize finite-energy GKP codewords with a general Fock-type envelope. We start by computing the normalized stabilizers of \cref{eq:norm_stab}. To compute the effect of the normalization on the stabilizers, we only need to understand how it acts on the quadrature operators since 
\begin{equation}\label{eq:damped_stab}
 E\pp{\Sigma_E,\mu_E} \hat{T}\left(\boldsymbol{v}\right) E\pp{\Sigma_E,\mu_E}^{-1} = {\rm e}^{-il \mathbf{v}^T \Omega \,E\pp{\Sigma_E,\mu_E}\,\boldsymbol{\hat{\xi}}\,E\pp{\Sigma_E,\mu_E}^{-1} }
\end{equation}
A general Gaussian envelope is described by a real symmetric matrix of covariance $\Sigma_E$ and a vector of means $\boldsymbol{\mu}_E$. Any real symmetric matrix has a symplectic decomposition $\Sigma_E = M\Sigma_0 M^T$, where $\Sigma_0=\varepsilon^{-1} \oplus \varepsilon^{-1}$ is diagonal and positive definite and $M$ is a symplectic matrix. Using this decomposition, the envelope operator can be rewritten as 
\begin{equation}
 E\pp{\Sigma_E,\mu_E} = \hat{T}\left(\boldsymbol{\mu}_E\right) U_{M}  E\pp{\Sigma_0,\boldsymbol{0}}U_{M}^\dagger\hat{T}\left(\boldsymbol{\mu}_E\right)^\dagger,
\end{equation}
which means that the action of the envelope on the quadrature operators can be written as
\begin{equation}
 E\pp{\Sigma_E,\mu_E}\,\boldsymbol{\hat{\xi}}\,E\pp{\Sigma_E,\mu_E}^{-1} = \hat{T}\left(\boldsymbol{\mu}_E\right) U_{M}  E\pp{\Sigma_0,\boldsymbol{0}} \bb{M\boldsymbol{\hat{\xi}} + \boldsymbol{\mu}_E} E\pp{\Sigma_0,\boldsymbol{0}}^{-1}U_{M}^\dagger\hat{T}\left(\boldsymbol{\mu}_E\right)^\dagger.
\end{equation}
The envelope operator $E\pp{\Sigma_0, \boldsymbol{0}}$ acts element-wise on each quadrature, leading to the simplification
\begin{equation}
 E\pp{\Sigma_E,\mu_E}\,\boldsymbol{\hat{\xi}}\,E\pp{\Sigma_E,\mu_E}^{-1} = \hat{T}\left(\boldsymbol{\mu}_E\right) U_{M} \bb{M E\pp{\Sigma_0,\boldsymbol{0}} \boldsymbol{\hat{\xi}}E\pp{\Sigma_0,\boldsymbol{0}}^{-1} + \boldsymbol{\mu}_E} U_{M}^\dagger\hat{T}\left(\boldsymbol{\mu}_E\right)^\dagger.
\end{equation}
For a diagonal covariance matrix $\Sigma_0$, we get 
\begin{equation}
  \begin{aligned}
  E\pp{\Sigma_0,\boldsymbol{0}} \boldsymbol{\hat{\xi}} E\pp{\Sigma_0,\boldsymbol{0}}^{-1} &= \bb{\cosh\pp{\Sigma_0^{-1}} + i\sinh\pp{\Sigma_0^{-1}}\Omega}\boldsymbol{\hat{\xi}},\\
  &\equiv h\pp{\Sigma_0}\boldsymbol{\hat{\xi}},
  \end{aligned}
\end{equation}
resulting in the full expression
\begin{equation}
 E\pp{\Sigma_E,\mu_E}\,\boldsymbol{\hat{\xi}}\,E\pp{\Sigma_E,\mu_E}^{-1} = M h\pp{\Sigma_0}M^{-1} \bb{\boldsymbol{\hat{\xi}}-\boldsymbol{\mu}_E }+  \boldsymbol{\mu}_E.
\end{equation}
Inserting this last equation into \cref{eq:damped_stab}, we arrive at the expression of the finite-energy code stabilizers
\begin{equation}
 \hat{T}_{\Sigma_E,\boldsymbol{\mu}_E}\pp{\boldsymbol{s}_j} = \exp\pp{-il \boldsymbol{s}_j^T\Omega\boldsymbol{\mu}_E}\exp\pp{-il\boldsymbol{s}_j^T\Omega Mh\pp{\Sigma_0}M^{-1}\bb{\boldsymbol{\hat{\xi}}-\boldsymbol{\mu}_E }}.
\end{equation}
A state described by a translated envelope is stabilized by the same protocol as a centered one, but with a different gauge. Moving forward, we will omit this phase for simplicity and focus solely on finding an implementation for the second term. Following the derivation of Ref.~\cite{royer_2020}, we compute the set of nullifiers
\begin{align}
 \hat{d}_j &= \frac{i}{\sqrt{2}} \ln\, \hat{T}_{\Sigma_E,\boldsymbol{\mu}_E}\pp{\boldsymbol{s}_j} \\ &=  \frac{l}{\sqrt{2}}\boldsymbol{s}_j^T \Omega M\cosh\pp{\Sigma_0^{-1}}M^{-1}\bb{\boldsymbol{\hat{\xi}}_\text{mod}-\boldsymbol{\mu}_E } +  \frac{il}{\sqrt{2}}\boldsymbol{s}_j^T \Omega M\sinh\pp{\Sigma_0^{-1}}\Omega M^{-1}\bb{\boldsymbol{\hat{\xi}}-\boldsymbol{\mu}_E }\nonumber
\end{align}
that enable the stabilization of the desired manifold given the implementation of the unitary 
\begin{equation}\label{eq:uni_sBs}
\hat  U_j = {\rm e}^{-i\Gamma_j \bb{\hat{d}_j \hat b^\dagger+ \hat{d}_j^\dagger \hat  b}}
\end{equation}
between the oscillators and a bath. The subscript on $\hat{\boldsymbol{\xi}}_\text{mod}$ means that we take the quadrature operators modulo a translation by $lM\sech\bb{\Sigma_0^{-1}}M^{-1}\boldsymbol{s}_j$, which comes from the branching of the logarithmic function. We choose to remove this condition on the quadrature operator, and instead apply this symmetry condition on the whole unitary $\hat{U}_j$. The cooling rate $\Gamma_j$ (dimensionless) is then fixed by this condition. Under a few assumptions (see Ref.~\cite{royer_2020}), we modelize the bath as a qubit with $b=\frac{\sigma_x+i\sigma_y}{\sqrt{2}}$ that is discarded after each interaction. We also use a second-order Trotterization on the oscillator-bath interaction. This leads to an approximation of the unitary in \cref{eq:uni_sBs}:
\begin{align}
\hat  U_j \approx &\exp\pp{ \frac{-il\Gamma_j}{2}\boldsymbol{s}_j^T \Omega M\sinh\pp{\Sigma_0^{-1}}\Omega M^{-1}\bb{\boldsymbol{\hat{\xi}}-\boldsymbol{\mu}_E } \sigma_y} \\ & \times \exp\pp{ -il\Gamma_j \boldsymbol{s}_j^T \Omega M\cosh\pp{\Sigma_0^{-1}}M^{-1}\bb{\boldsymbol{\hat{\xi}}-\boldsymbol{\mu}_E } \sigma_x}\\  &\times \exp\pp{ \frac{-il\Gamma_j}{2}\boldsymbol{s}_j^T \Omega M\sinh\pp{\Sigma_0^{-1}}\Omega M^{-1}\bb{\boldsymbol{\hat{\xi}}-\boldsymbol{\mu}_E } \sigma_y}. \nonumber
\end{align}  
Asking for this unitary to be symmetric under the translation mentioned before results in the expression
\begin{equation}
  \Gamma_j = \pp{\boldsymbol{s}_j^T M^{-T} \tanh\pp{\Sigma_0^{-1}} M^{-1}\boldsymbol{s}_j }^{-1}
\end{equation}
for the cooling rate. This unitary can be cast in a circuit that is implemented only with controlled displacements and qubit rotations
\begin{equation}
 \text{C}\hat{T}\pp{\boldsymbol{v}} = \exp\pp{-il\boldsymbol{v}^T\Omega \hat{\boldsymbol{\xi}}\frac{\hat{\sigma}_z}{2}},\qquad R_{x}\pp{\theta} = \exp\pp{-i\theta\frac{\hat{\sigma}_x}{2}}.
\end{equation}
In this representation, we obtain the unitary operator
\begin{align}
  \hat{U}_j \approx \, & \text{C}\hat{T}\pp{\boldsymbol{\alpha}_j}\hat{R}_z\pp{l\boldsymbol{\alpha}_j^T \Omega \boldsymbol{\mu}_E}\hat{R}_x^\dagger\pp{\frac{\pi}{2}}\hat{R}_z\pp{-l\boldsymbol{\beta}_j^T \Omega \boldsymbol{\mu}_E}\text{C}\hat{T}\pp{\boldsymbol{\beta}_j} \\
  & \times \hat{R}_z\pp{-l\boldsymbol{\beta}_j^T \Omega \boldsymbol{\mu}_E} \hat{R}_x\pp{\frac{\pi}{2}} \hat{R}_z\pp{l\boldsymbol{\alpha}_j^T \Omega \boldsymbol{\mu}_E}\text{C}\hat{T}\pp{\boldsymbol{\alpha}_j},
\end{align}
with small and big controlled displacements $\alpha_j$ and $\beta_j$, respectively defined as
\begin{equation}
  \boldsymbol{\alpha}_j = -\frac{ M\sinh\pp{\Sigma_0^{-1}}M^T\Omega \boldsymbol{s}_j}{\boldsymbol{s}_j^TM^{-T}\tanh\pp{\Sigma_0^{-1}}M^{-1}\boldsymbol{s}_j}, \quad  \boldsymbol{\beta}_j = \frac{2 M\cosh\pp{\Sigma_0^{-1}}M^{-1} \boldsymbol{s}_j}{\boldsymbol{s}_j^TM^{-T}\tanh\pp{\Sigma_0^{-1}}M^{-1}\boldsymbol{s}_j}.
\end{equation}
This is the circuit presented in \cref{fig:sBs_circ} of \cref{sec:cct}. A major difference with the sBs circuit of \cite{royer_2020} is that the small and big displacements are not orthogonal for a general symplectic matrix $M$. Another difference is that, for non-zero $\boldsymbol{\mu}_E$, we get additional correction to the qubit rotations. Those rotations can be combined with the original $\pi/2$ rotations around the $\sigma_{x}$ axis to keep the number of operations constant.

\section{Wigner function of GKP codes projectors}\label{App_Wig_proj}
In this section, we derive the Wigner representation of the codespace projector of an arbitrary GKP code considering a gauge. This demonstration is inspired from appendix 
A in \cite{conrad_2022}. Our approach complete the demonstration for non-trivial gauges and non symplectically even lattice. As defined in \cite{conrad_2022,royer_2022}, an element of the grid $\boldsymbol{\lambda}\in \Lambda$ of \cref{eq:stab} with generator matrix $S$ is related to an element of the stabilizer group $\mathcal{S}$ up to a gauge $\boldsymbol{\mu}\in\mathbb{Z}_{2}^{2N}$ by the stabilizer equation:
\begin{equation}
 \mathcal{S}= \left\{\nu_{\boldsymbol{\mu}}\pp{\boldsymbol{\lambda}}\hat{T}\pp{\boldsymbol{\lambda}}\ket{\psi} = \ket{\psi},\,\forall \boldsymbol{\lambda}\in \Lambda\right\},
\end{equation} 
where 
\begin{equation}
  \nu_{\boldsymbol{\mu}}\pp{\boldsymbol{\lambda}} = \exp\pp{i\pi \boldsymbol{\lambda}^T S^{-1}\bb{A_\text{\lltriangle} S^{-T}\boldsymbol{\lambda}+\boldsymbol{\mu}}},
\end{equation}
and $A_\text{\lltriangle}$ is the lower triangular part of the symplectic Gram matrix $A=S\Omega S^T$. After moving to the canonical basis $S_D = RS$ using the matrix $R$ in \cref{eq:canonical}, we have a simplified equation for the gauge
\begin{equation}
  \nu_{\boldsymbol{\mu}}\pp{\boldsymbol{\lambda}} = \exp\pp{i\pi \boldsymbol{a}^T \bb{A_{D\text{\lltriangle}} \boldsymbol{a}+R\boldsymbol{\mu}}},
\end{equation}
where we used $\boldsymbol{a} = S_D^{-T}\boldsymbol{\lambda} \in \mathbb{Z}^{2N}$. This transformation does not change the properties of the code, and we obtain the codespace projector as
\begin{equation}
  \Pi_{S} = \sum_{\boldsymbol{\lambda}\in \Lambda}\nu_{\boldsymbol{\mu}}\pp{\boldsymbol{\lambda}}\hat{T}\pp{\boldsymbol{\lambda}} 
\end{equation}
The Wigner representation of this projector is then computed using the definition in \cref{eq:wigner_def}
\begin{equation}
 W\pp{\boldsymbol{\xi},\Pi_S} = \int_{\mathbb{R}^{2N}}d\boldsymbol{v}\, e^{-i{2\pi}\boldsymbol{\xi}^T\Omega \boldsymbol{v}}\, \text{Tr}\bb{\hat{T}\pp{\boldsymbol{v}}\Pi_S},
\end{equation}
\begin{equation}
 W\pp{\boldsymbol{\xi},\Pi_S} = \sum_{\boldsymbol{\lambda}\in \Lambda}\nu_{\boldsymbol{\mu}}\pp{\boldsymbol{\lambda}}\int_{\mathbb{R}^{2N}}d\boldsymbol{v}\, e^{-i{2\pi}\boldsymbol{\xi}^T\Omega \boldsymbol{v}}\, \text{Tr}\bb{\hat{T}\pp{\boldsymbol{v}}\hat{T}\pp{\boldsymbol{\lambda}}}.
\end{equation}
Since translation operators are trace orthogonal: $\text{Tr}\bb{\hat{T}\pp{\boldsymbol{v}}\hat{T}\pp{\boldsymbol{\lambda}}} = \delta^{2N}\pp{\boldsymbol{v}+\boldsymbol{\lambda}}$, we obtain 
\begin{equation}
 W\pp{\boldsymbol{\xi},\Pi_S} = \sum_{\boldsymbol{\lambda}\in \Lambda}\nu_{\boldsymbol{\mu}}\pp{\boldsymbol{\lambda}}\, e^{i{2\pi}\boldsymbol{\xi}^T\Omega \boldsymbol{\lambda}}.
\end{equation}
\begin{equation}\label{eq:theta_sum}
 W\pp{\boldsymbol{\xi},\Pi_S} = \sum_{\boldsymbol{a} \in \mathbb{Z}^{2N}}\exp\pp{\pi i \boldsymbol{a}^T A_{D\text{\lltriangle}} \boldsymbol{a}+2\pi i \bb{\frac{R}{2}\boldsymbol{\mu} -S_D\Omega \boldsymbol{\xi}}^T \boldsymbol{a}},
\end{equation} 
which can be rewritten in terms of a multivariable theta function \cite{mumford_2007a}
\begin{equation}
 W\pp{\boldsymbol{\xi},\Pi_S} = \frac{1}{\text{det}\pp{\Lambda}}\vartheta\pp{\frac{R}{2}\boldsymbol{\mu} -S_D\Omega \boldsymbol{\xi}; A_{D\text{\lltriangle}}}.
\end{equation} 
We can go further in the description using the symmetry of the theta function in the second argument. Reminding ourselves that $A_D = \Omega_2 \otimes D$, only the components for which $D_{kk}$ is odd contribute to a non-trivial phase. Defining the matrix $D_2 = \pp{D\oplus D} \text{ mod }2$, we can split the infinite sum into two part on $\boldsymbol{b}\in\mathbb{Z}^{2N}$ and $\boldsymbol{m}\in\mathbb{Z}_2^{2N}$:
\begin{equation}
 \boldsymbol{a} = \pp{I_{2N}+D_2}\boldsymbol{b} + D_2\boldsymbol{m}.
\end{equation}
This manipulation separates the even and odd components of the vector $\boldsymbol{a}$, but only on the part for which the lattice is not symplectically even : 
\begin{equation}
 \boldsymbol{a}_k = 
     \begin{cases}
 \boldsymbol{b}_k &\quad\text{if }D_{kk}\text{ mod }\,2=0 \\
      2\boldsymbol{b}_k+\boldsymbol{m}_k &\quad\text{if }D_{kk}\text{ mod }\,2=1 \\
     \end{cases}
\end{equation}
Using this change of summation indexing, we find that the products $\pp{I_{2N}+D_2} A_{D\text{\lltriangle}} = A_{D\text{\lltriangle}} \pp{I_{2N}+D_2}$ is an even matrix in $2\mathbb{Z}^{2N\times 2N}$. This simplify the Wigner function of \cref{eq:theta_sum} to
\begin{equation}
  \sum_{\boldsymbol{m} \in \mathbb{Z}_{2}^{2N}}e^{\pp{\pi i \boldsymbol{m}^T D_2 A_{D\text{\lltriangle}}D_2 \boldsymbol{m}+2\pi i \bb{\frac{D_2R}{2}\boldsymbol{\mu} -D_2S_D\Omega \boldsymbol{\xi}}^T \boldsymbol{m}}}\sum_{\boldsymbol{b} \in \mathbb{Z}^{2N}}e^{2\pi i \bb{\frac{\pp{I+D_2}R}{2}\boldsymbol{\mu}-\pp{I+D_2}S_D\Omega\boldsymbol{\xi}}^T\boldsymbol{b}}.
\end{equation}
The first sum in $\boldsymbol{m}$, that we note $C\pp{S, \boldsymbol{\mu}, \boldsymbol{\xi}}$, is finite and can be computed for a specified lattice. For the second series, we can compute it as a product of Poisson summations to find
\begin{equation}
  \sum_{\boldsymbol{b} \in \mathbb{Z}^{2N}}e^{2\pi i \bb{\frac{\pp{I+D_2}R}{2}\boldsymbol{\mu}-\pp{I+D_2}S_D\Omega\boldsymbol{\xi}}^T\boldsymbol{b}} = \frac{1}{\det\pp{\tilde{\Lambda}}} \sum_{\tilde{\boldsymbol{\lambda}}^*\in \tilde{\Lambda}^* } \delta^{\pp{2N}}\pp{\boldsymbol{\xi}-\bb{\tilde{\boldsymbol{\lambda}}^*-\frac{1}{2}S^* \boldsymbol{\mu} }}
\end{equation}
with $S^* = \Omega S^{-1}$. We have also defined a partially rescale lattice $\tilde{\Lambda}$:
\begin{equation}
 \tilde{\Lambda} = \left\{ \bb{\pp{I+D_2}S_D}^T\boldsymbol{a}\, |\, \boldsymbol{a}\in\mathbb{Z}^{2N}    \right\},
\end{equation}
and its symplectic dual lattice 
\begin{equation}
 \tilde{\Lambda}^* = \left\{ \Omega\bb{\pp{I+D_2}S_D}^{-1}\boldsymbol{a}\, |\, \boldsymbol{a}\in\mathbb{Z}^{2N}    \right\},
\end{equation}
We finally arrive at the result for the Wigner function for the projector onto the codespace of a GKP code with generator matrix $S$ and gauge $\boldsymbol{\mu}$:
\begin{equation}\label{eq:Wign_prop_final}
 W\pp{\boldsymbol{\xi},\Pi_S} = \frac{1}{\det\pp{\tilde{\Lambda}}}C\pp{S, \boldsymbol{\mu}, \boldsymbol{\xi}}\sum_{\tilde{\boldsymbol{\lambda}}^*\in \tilde{\Lambda}^* } \delta^{\pp{2N}}\pp{\boldsymbol{\xi}-\bb{\tilde{\boldsymbol{\lambda}}^*-\frac{1}{2}S^* \boldsymbol{\mu} }}.
\end{equation}
For the final result of this section, we produce the Wigner function of a pure grid state $\ket{\Psi_S}$. Provided that we are looking at the projector over a pure GKP grid state $\ket{\Psi_S}$. Then, the Wigner function of the projector on $\ket{\Psi_S}$ is the Wigner function of the state itself. In this case, because the canonical symplectic Gram matrix is only $\Omega$, $D_2=I$ and \cref{eq:Wign_prop_final} simplify to 
\begin{equation}\label{eq:Wign_state_final}
 W\pp{\boldsymbol{\xi},\ket{\Psi_S}} = \frac{C\pp{S, \boldsymbol{\mu}, \boldsymbol{\xi}}}{4^{2N}}\sum_{{\boldsymbol{\lambda}}^*\in {\Lambda}^* } \delta^{\pp{2N}}\pp{\boldsymbol{\xi}-\frac{1}{2}\bb{{\boldsymbol{\lambda}}^*-S^* \boldsymbol{\mu} }},
\end{equation}
with $\tilde{S}=2S$. As a result, all grid states can be represented in phase-space as a sum of Dirac delta functions positioned at half the dual lattice vectors under a trivial gauge ($\boldsymbol{\mu} = \boldsymbol{0}$):
\begin{equation}
 W\pp{\boldsymbol{\xi},\ket{\Psi_S}} \propto \sum_{{\boldsymbol{\lambda}}^*\in {\Lambda}^* } C\pp{S, \boldsymbol{0}, \frac{\boldsymbol{\lambda}^*}{2} } \delta^{\pp{2N}}\pp{\boldsymbol{\xi}- \frac{\boldsymbol{\lambda}^*}{2}}.
\end{equation}

\section{Wigner function of multimode finite energy GKP codes states}\label{App_Wig_FE_states}
As described in the main text, a general pure grid state of a finite energy GKP code is described by the infinite energy grid state and an envelope $E\pp{\Sigma_E, \boldsymbol{\mu}_E}$:
\begin{equation}
\ket{\Psi_{S,\Sigma_E,\boldsymbol{\mu}_E}} = \hat{E}\pp{\Sigma_E,\boldsymbol{\mu}_E}\ket{\Psi_{S}}.
\end{equation}
In this section, we derive the Wigner representation of this finite energy grid state as a function of $S$, $\Sigma_E$ and $\boldsymbol{\mu}_E$. The Wigner representation of this state has a simple decomposition as a star-product of Wigner functions of the envelope and the infinite energy state
\begin{equation}\label{eq:wig_state}
 W\pp{\boldsymbol{\xi},\ket{\Psi_{S,\Sigma_E,\boldsymbol{\mu}_E}}} = W\pp{\boldsymbol{\xi},\hat{E}} \star W\pp{\boldsymbol{\xi},\ket{\Psi_S}} \star W\pp{\boldsymbol{\xi},\hat{E}}.
\end{equation}
This star-product is defined as 
\begin{equation}
 A\star B :=A\, \prod_{j=1}^{N}\exp\pp{\frac{i}{2} \bb{\cev{\partial}_{q_{j}}\vec{\partial}_{p_{j}}-\cev{\partial}_{p_{j}}\vec{\partial}_{q_{j}}  }} \,B
\end{equation}
with $\vec{\partial}_a$ the partial derivative with respect to $a$ acting in the direction of the arrow. Another equivalent definition for the star-product is \cite{baker_1958}
\begin{equation}
 A\pp{\boldsymbol{\xi}}\star B\pp{\boldsymbol{\xi}} = 2^{2N} \int_{\mathbb{R}^{2N}}d\boldsymbol{\eta}d\boldsymbol{\nu} A\pp{\boldsymbol{\xi}+\boldsymbol{\eta}} B\pp{\boldsymbol{\xi}+\boldsymbol{\nu}} \mathrm{e}^{4\pi i\boldsymbol{\eta}^T\Omega \boldsymbol{\nu}},
\end{equation}
which has been rescaled according to our definition of the Wigner function in \cref{eq:wigner_def}.
This product is associative and bilinear. Combining the argument of bilinearity with the result of the last section that a grid state has a Wigner function that decomposes as a sum of Dirac delta functions (See \cref{App_Wig_proj}), \cref{eq:wig_state} reduces to
\begin{equation}\label{eq:GKP_state_wigner}
 W\pp{\boldsymbol{\xi},\ket{\Psi_{S,\Sigma_E,\boldsymbol{\mu}_E}}} =\sum_{{\boldsymbol{\lambda}}^*\in {\Lambda}^* } c_{\boldsymbol{\lambda}^*}\, W\pp{\boldsymbol{\xi},\hat{E}} \star \delta^{\pp{2N}}\pp{\boldsymbol{\xi}-\frac{1}{2}\bb{{\boldsymbol{\lambda}}^*-S^* \boldsymbol{\mu} }} \star W\pp{\boldsymbol{\xi},\hat{E}}.
\end{equation}
where $c_{\boldsymbol{\lambda}^*} = C\pp{S, \boldsymbol{\mu}, \frac{\boldsymbol{\lambda}^*}{2}}/4^{2N}$. Then, we only need to compute each term of the sum:
\begin{equation}\label{eq:env_star_prod}
 W\pp{\boldsymbol{\xi},\hat{E}} \star \delta^{\pp{2N}}\pp{\boldsymbol{\xi}-{\boldsymbol{\chi}}} \star W\pp{\boldsymbol{\xi},\hat{E}}.
\end{equation} 
To achieve this, we first compute the Wigner representation of the envelope. A general envelope $\hat{E}\pp{\Sigma_E,\boldsymbol{\mu}_E}$ can always be decomposed as a displaced and squeezed Gaussian function 
\begin{equation}
 \hat{E}\pp{\Sigma_E,\boldsymbol{\mu}_E} = \hat{E}\pp{M_E\Sigma_0 M_E^T,\boldsymbol{\mu}_E} = \hat{T}\left(\boldsymbol{\mu}_E\right)  U_{M_E} \hat{E}\pp{\Sigma_0 ,\boldsymbol{0}} U_{M_E}^\dagger \hat{T}\left(\boldsymbol{\mu}_E\right)^\dagger,
\end{equation}
where $\Sigma_0$ is the symplectic diagonalisation of $\Sigma_E$. This canonical form of the envelope is connected to a thermal state of temperature $\Sigma_0^{-1}$ in multiple modes. They are explicitly related as
\begin{equation}
\hat{E}\pp{\Sigma_0 ,\boldsymbol{0}}= \frac{1}{\sqrt{\det\pp{I-\exp\pp{-\Sigma_0^{-1}}}}} \hat{\rho}_{\text{thermal}}.
\end{equation}
The full thermal state is the product of one-mode thermal states. Therefore, $\hat{\rho}_{\text{thermal}}$ is a product of Gaussian states with zero mean and covariance matrix $\pp{2\overline{n}_k+1}I/4\pi$ \cite{weedbrook_2012}, where $\overline{n}_k$ is the mean number of excitation in mode $k$. Alternatively, we can use the connection $\overline{n}_k=\bb{\exp\pp{\varepsilon_k}}^{-1}$ between excitation number and temperature of the thermal state to find that its covariance matrix can also be written as $\coth\pp{\varepsilon_k/2}I/4\pi$. Combining $N$ single-mode thermal states result in a Gaussian function with a covariance matrix
\begin{equation}
  \frac{1}{4\pi} \coth\pp{\Sigma_0^{-1}/2}.
\end{equation}
Using this connection with the thermal state, we find that the Wigner representation of this simple envelope is given by
\begin{equation}
 W\pp{\boldsymbol{\xi};\hat{E}\pp{\Sigma_0,\boldsymbol{0}}} = \det\pp{I-\exp\pp{-\Sigma_0^{-1}}}^{-1/2}\, G_{\frac{1}{4\pi}\coth\pp{\Sigma_0^{-1}/2},\boldsymbol{0}}\pp{\boldsymbol{\xi}},
\end{equation}
with $G$ defined in \cref{eq:gauss_wig}. From this result, we can obtain the Wigner representation of a general envelope by using \cref{eq:mod_wigner}:
\begin{equation}
 W\pp{\boldsymbol{\xi};\hat{E}\pp{\Sigma_E,\boldsymbol{\mu}_E}} = \det\pp{I-\exp\pp{-\Sigma_0^{-1}}}^{-1/2}\, G_{\frac{1}{4\pi}M_E\coth\pp{\Sigma_0^{-1}/2}M_E^T,\boldsymbol{\mu}_E}\pp{\boldsymbol{\xi}}.
\end{equation}
To simplify the notation, we rewrite this last equation as 
\begin{equation}
 W\pp{\boldsymbol{\xi};\hat{E}\pp{\Sigma_E,\boldsymbol{\mu}_E}} = \mathcal{N}\, G_{A,\boldsymbol{a}}\pp{\boldsymbol{\xi}},
\end{equation}
where $A=\frac{1}{4\pi}M_E\coth\pp{\Sigma_0^{-1}/2}M_E^T$ and $\boldsymbol{a} = \boldsymbol{\mu}_E$. With this intermediate result, we can move on to compute \cref{eq:env_star_prod}. Starting with the left star-product, we have 
\begin{equation}\label{eq:Frist_starprod}
 W\pp{\boldsymbol{\xi},\hat{E}} \star \delta^{\pp{2N}}\pp{\boldsymbol{\xi}-{\boldsymbol{\chi}}}  = 2^{2N}\mathcal{N} \int_{\mathbb{R}^{2N}}d\boldsymbol{\eta}d\boldsymbol{\nu}\, G_{A,\boldsymbol{a}}\pp{\boldsymbol{\xi}+\boldsymbol{\eta}} \delta^{\pp{2N}}\pp{\boldsymbol{\xi}+\boldsymbol{\nu}-\boldsymbol{\chi}} \mathrm{e}^{4\pi i\boldsymbol{\eta}^T\Omega \boldsymbol{\nu}}. 
\end{equation}
We then use a change of variable $\tilde{\boldsymbol{\eta}} = \boldsymbol{\eta} + \boldsymbol{\xi}$ 
\begin{equation}
 (\ref{eq:Frist_starprod}) = 2^{2N}\mathcal{N} \int_{\mathbb{R}^{2N}}\,d\boldsymbol{\nu}\,\delta^{\pp{2N}}\pp{\boldsymbol{\xi}+\boldsymbol{\nu}-\boldsymbol{\chi}} \mathrm{e}^{-4\pi i\boldsymbol{\xi}^T\Omega {\boldsymbol{\nu}}}\int_{\mathbb{R}^{2N}}d\tilde{\boldsymbol{\eta}}\, G_{A,\boldsymbol{a}}\pp{\tilde{\boldsymbol{\eta}}} \mathrm{e}^{4\pi i\tilde{\boldsymbol{\eta}}^T\Omega {\boldsymbol{\nu}}}. 
\end{equation}
The integral over $\tilde{\boldsymbol{\eta}}$ can be identified as the characteristic function of a Gaussian distribution at the position $4\pi\Omega \nu$ to obtain
\begin{equation}
 (\ref{eq:Frist_starprod}) = 2^{2N}\mathcal{N} \int_{\mathbb{R}^{2N}}\,d\boldsymbol{\nu}\,\delta^{\pp{2N}}\pp{\boldsymbol{\xi}+\boldsymbol{\nu}-\boldsymbol{\chi}} \mathrm{e}^{4\pi i \bb{\boldsymbol{a}-\boldsymbol{\xi}}^T\Omega {\boldsymbol{\nu}}} \exp\pp{\frac{-16\pi^2}{2}\boldsymbol{\nu}^T\Omega A \Omega^T \boldsymbol{\nu}}.
\end{equation}
We complete the computation of the product using the Dirac delta distribution inside the integral 
\begin{equation}
 (\ref{eq:Frist_starprod}) = 2^{2N}\mathcal{N} \mathrm{e}^{4\pi i \bb{\boldsymbol{\xi}-\boldsymbol{a}}^T\Omega \bb{\boldsymbol{\xi}-\boldsymbol{\chi}}} \exp\pp{\frac{-16\pi^2}{2}\bb{\boldsymbol{\xi}-\boldsymbol{\chi}}^T\Omega A \Omega^T \bb{\boldsymbol{\xi}-\boldsymbol{\chi}}}.
\end{equation}
With the first star product computed, we now proceed to the second part 
\begin{equation}
\begin{split}
(\ref{eq:env_star_prod}) = &2^{2N}\mathcal{N}^2 \int_{\mathbb{R}^{2N}}d\boldsymbol{\eta}d\boldsymbol{\nu}\, \mathrm{e}^{4\pi i \bb{\boldsymbol{\xi}+\boldsymbol{\eta}-\boldsymbol{a}}^T\Omega \bb{\boldsymbol{\xi}+\boldsymbol{\eta}-\boldsymbol{\chi}}}\,\times \\ & \exp\pp{\frac{-16\pi^2}{2}\bb{\boldsymbol{\xi}+\boldsymbol{\eta}-\boldsymbol{\chi}}^T\Omega A \Omega^T \bb{\boldsymbol{\xi}+\boldsymbol{\eta}-\boldsymbol{\chi}}}
 G_{A,\boldsymbol{a}}\pp{\boldsymbol{\xi}+\boldsymbol{\nu}}  \mathrm{e}^{4\pi i\boldsymbol{\eta}^T\Omega \boldsymbol{\nu}}.
\end{split}
\end{equation}
Once again, we use a change of variable $\tilde{\boldsymbol{\nu}} = \boldsymbol{\nu} + \boldsymbol{\xi}$ and identify the characteristic function of a Gaussian distribution to compute a first integral. Combining two Gaussian functions and using the property $\omega\pp{\boldsymbol{\nu},\boldsymbol{\nu}}=0$ to simplify phase terms, we arrive at \cref{eq:3of4_int}
\begin{equation}\label{eq:3of4_int}
\begin{split}
(\ref{eq:env_star_prod}) = &2^{2N}\mathcal{N}^2 \exp\pp{\frac{-8\pi^2}{2}\bb{\boldsymbol{\xi}-\boldsymbol{\chi}}^T\Omega A \Omega^T \bb{\boldsymbol{\xi}-\boldsymbol{\chi}}} \mathrm{e}^{4\pi i \bb{\boldsymbol{\xi}-\boldsymbol{a}}^T\Omega \bb{\boldsymbol{\xi}-\boldsymbol{\chi}}} \,\times\\ &\int_{\mathbb{R}^{2N}}d\boldsymbol{\eta}\, 
\mathrm{e}^{4\pi i \bb{\boldsymbol{\xi}+\boldsymbol{\chi}-2\boldsymbol{a}}^T\Omega \boldsymbol{\eta}}
\exp\pp{\frac{-32\pi^2}{2}\bb{\boldsymbol{\eta}+\frac{\boldsymbol{\xi}-\boldsymbol{\chi}}{2}}^T\Omega A \Omega^T \bb{\boldsymbol{\eta}+\frac{\boldsymbol{\xi}-\boldsymbol{\chi}}{2}}}.
\end{split}
\end{equation}
The last exponential function in \cref{eq:3of4_int} is linked to a Gaussian distribution function through
\begin{equation}
 \exp\pp{\frac{-32\pi^2}{2}\bb{\boldsymbol{\eta}+\frac{\boldsymbol{\xi}-\boldsymbol{\chi}}{2}}^T\Omega A \Omega^T \bb{\boldsymbol{\eta}+\frac{\boldsymbol{\xi}-\boldsymbol{\chi}}{2}}} = \sqrt{\text{det}\pp{\frac{A^{-1}}{16\pi}}} G_{\frac{\Omega A^{-1} \Omega^T}{32\pi^2},\frac{\boldsymbol{\chi}-\boldsymbol{\xi}}{2}}\pp{\boldsymbol{\eta}},  
\end{equation}
with 
\begin{equation}
 \text{det}\pp{\frac{A^{-1}}{16\pi}} = 2^{-4N}\text{det}\pp{\tanh\pp{\Sigma_0^{-1}/2}},
\end{equation}
once we insert back the initial form of $A$. With this identification, we can again use the definition of the characteristic function of a Gaussian distribution to compute this last integral
\begin{equation}\label{eq:4of4_int}
\begin{split}
(\ref{eq:env_star_prod}) = &\sqrt{\text{det}\pp{\tanh\pp{\Sigma_0^{-1}/2}}}\mathcal{N}^2 \exp\pp{\frac{-8\pi^2}{2}\bb{\boldsymbol{\xi}-\boldsymbol{\chi}}^T\Omega A \Omega^T \bb{\boldsymbol{\xi}-\boldsymbol{\chi}}} \mathrm{e}^{4\pi i \bb{\boldsymbol{\xi}-\boldsymbol{a}}^T\Omega \bb{\boldsymbol{\xi}-\boldsymbol{\chi}}}\times \\ &\int_{\mathbb{R}^{2N}}d\boldsymbol{\eta}\, 
\mathrm{e}^{ i \pp{-4\pi\Omega\bb{\boldsymbol{\xi}+\boldsymbol{\chi}-2\boldsymbol{a}}}^T \boldsymbol{\eta}}
G_{\frac{\Omega A^{-1} \Omega^T}{32\pi^2},\frac{\boldsymbol{\chi}-\boldsymbol{\xi}}{2}}\pp{\boldsymbol{\eta}}.
\end{split}
\end{equation}
After a few manipulations, we get the simplified equation
\begin{equation}\label{eq:final_star_product_result}
\begin{split}
(\ref{eq:env_star_prod}) = &\sqrt{\text{det}\pp{\tanh\pp{\Sigma_0^{-1}/2}}}\mathcal{N}^2 \exp\pp{\frac{-8\pi^2}{2}\bb{\boldsymbol{\xi}-\boldsymbol{\chi}}^T\Omega A \Omega^T \bb{\boldsymbol{\xi}-\boldsymbol{\chi}}}\times \\ &
 \exp\pp{\frac{-1}{2}\bb{\boldsymbol{\xi}+\boldsymbol{\chi}-2\boldsymbol{a}}^T\frac{A^{-1}}{2}\bb{\boldsymbol{\xi}+\boldsymbol{\chi}-2\boldsymbol{a}}}.
\end{split}
\end{equation}
At this point, it is useful to combine both exponentials into a single Gaussian function by completing the square in the exponent. After inserting the definition of $A$ and $\boldsymbol{a}$ into \cref{eq:final_star_product_result}, we get 
\begin{equation}\label{eq:final_result}
\begin{split}
(\ref{eq:env_star_prod}) = &\frac{\sqrt{\text{det}\pp{\tanh\pp{\Sigma_0^{-1}/2}}}}{\text{det}\pp{I-\exp\pp{-\Sigma_0^{-1}}}} \exp\pp{\frac{-1}{2}\bb{\boldsymbol{\chi}-\boldsymbol{\mu}_E}^T \bb{\frac{M_E \coth\pp{\Sigma_0^{-1}}M_E^T}{4\pi}}^{-1} \bb{\boldsymbol{\chi}-\boldsymbol{\mu}_E}} \, \times \\ &
\exp\pp{\frac{-1}{2}\bb{\boldsymbol{\xi}-\boldsymbol{\chi}_E}^T \bb{\frac{M_E \tanh\pp{\Sigma_0^{-1}}M_E^T}{4\pi}}^{-1}
 \bb{\boldsymbol{\xi}-\boldsymbol{\chi}_E}},
\end{split}
\end{equation}
where we have extracted the altered peak position 
\begin{equation}\label{eq:FE_chis}
 \boldsymbol{\chi}_E = M_E\sech\pp{\Sigma_0^{-1}}M_E^{-1}\bb{\boldsymbol{\chi}-\boldsymbol{\mu}_E} + \boldsymbol{\mu}_E.
\end{equation}
\Cref{eq:final_result} tells us that the normalization process of adding a Fock-type envelope to a GKP state Wigner function maps each Dirac distribution of the sum in \cref{eq:GKP_state_wigner} to a Gaussian function of amplitude described by the first line of \cref{eq:final_result}, centered around $\boldsymbol{\chi}_E$ and with covariance matrix $M_E \tanh\pp{\Sigma_0^{-1}}M_E^T/4\pi$. This describes entirely the Wigner function of any finite-energy GKP state with a Fock-type envelope. Especially, \cref{eq:final_result} reduces to the result of Ref.~\cite{bourassa_2021} when $M_E = I$ and $\boldsymbol{\mu}_E=\boldsymbol{0}$. If we set $\Sigma_0 = \alpha^{-1} I$ and take the limit $\alpha\to 0^+$, we recover the infinite energy state. In this limit, the peaks position are unaltered $\boldsymbol{\chi}_E \to \boldsymbol{\chi}$, each peak is infinitely squeezed with an infinite amplitude because $\tanh\pp{\Sigma_0}\to 0$ and the envelope becomes infinitely large because $\coth\pp{\Sigma_0}\to \infty$. 

\section{Average number of excitation in GKP states}\label{App_Wig_number}
In this section, we aim to obtain an expression for the average number of excitations in a GKP state with generator matrix $S$ and an envelope described by covariance matrix $\Sigma_E$ and position in phase-space $\boldsymbol{\mu}_E$. In \cref{App_Wig_FE_states}, we have derived the Wigner function of such a state, and in the following, we are going to use this result. From our definition of the Wigner function
\begin{equation}\label{eq:app_wigner_def}
 W\pp{\boldsymbol{\xi},\hat{\rho}} = \int_{\mathbb{R}^{2N}}d\boldsymbol{v}\, \mathrm{e}^{-i{2\pi}\boldsymbol{\xi}^T\Omega \boldsymbol{v}}\, \text{Tr}\bb{\hat{T}\pp{\boldsymbol{v}}\hat{\rho}},
\end{equation}
we can derive the expression for the average excitation number \cite{carmichael_1999}
\begin{equation}
 \left<\hat{\boldsymbol{\xi}}^T\hat{\boldsymbol{\xi}}\right> = \text{Tr}\bb{\hat{\boldsymbol{\xi}}^T\hat{\boldsymbol{\xi}}\hat{\rho}} = \pp{2\pi}^{2N}\int_{\mathbb{R}^{2N}}d\boldsymbol{\xi}\, \boldsymbol{\xi}^T\boldsymbol{\xi}W\pp{\boldsymbol{\xi},\hat{\rho}},
\end{equation}
for a state described by the Wigner function $W\pp{\boldsymbol{\xi},\hat{\rho}}$. With this intermediate result, we are ready to compute it with the Wigner function of GKP states from \cref{eq:final_result}. To simplify the demonstration, we express the Wigner function of GKP states as 
\begin{equation}
 W\pp{\boldsymbol{\xi},\hat{\rho}_{\text{GKP}}} = \sum_{\boldsymbol{\chi}\in\mathcal{M}} c_{\boldsymbol{\chi}} \exp\pp{\frac{-1}{2}\bb{\boldsymbol{\xi}-\boldsymbol{\chi}_E}^T B^{-1}
 \bb{\boldsymbol{\xi}-\boldsymbol{\chi}_E}},
\end{equation}
with $\mathcal{M} = \pp{\Lambda^*-S^{*}\boldsymbol{\mu}}/2$ from \cref{eq:Wign_state_final}, $\boldsymbol{\chi_E}$ from \cref{eq:FE_chis} and \newline $B = M_E \tanh\pp{\Sigma_0^{-1}}M_E^T/4\pi$. This leads us to compute the integral 
\begin{equation}
 \left<\hat{\boldsymbol{\xi}}^T\hat{\boldsymbol{\xi}}\right> = \pp{2\pi}^{2N}\sum_{\boldsymbol{\chi}\in\mathcal{M}} c_{\boldsymbol{\chi}} \int_{\mathbb{R}^{2N}}d\boldsymbol{\xi}\, \boldsymbol{\xi}^T\boldsymbol{\xi} \exp\pp{\frac{-1}{2}\bb{\boldsymbol{\xi}-\boldsymbol{\chi}_E}^T B^{-1}
 \bb{\boldsymbol{\xi}-\boldsymbol{\chi}_E}}.
\end{equation}
This is an integral that we can compute analytically using a change of variable and adding a source term. We reframe the problem as solving the integral 
\begin{equation}
 \left. \int_{\mathbb{R}^{2N}}d\boldsymbol{\xi}\, \bb{\boldsymbol{\xi}+\boldsymbol{\chi}_E}^T\bb{\boldsymbol{\xi}+\boldsymbol{\chi}_E} \exp\pp{\frac{-1}{2}\boldsymbol{\xi}^T B^{-1}
 \boldsymbol{\xi} + \boldsymbol{J}^T\boldsymbol{\xi}} \right|_{\boldsymbol{J}=\boldsymbol{0}}.
\end{equation}
With this trick, we can move the polynomial term out of the integral by changing the variable for derivatives in $\boldsymbol{J}$
\begin{equation}
 \left. \bb{\frac{\partial}{\partial \boldsymbol{J}}+\boldsymbol{\chi}_E}^T\bb{\frac{\partial}{\partial \boldsymbol{J}}+\boldsymbol{\chi}_E} \int_{\mathbb{R}^{2N}}d\boldsymbol{\xi}\,\exp\pp{\frac{-1}{2}\boldsymbol{\xi}^T B^{-1}
 \boldsymbol{\xi} + \boldsymbol{J}^T\boldsymbol{\xi}} \right|_{\boldsymbol{J}=\boldsymbol{0}},
\end{equation}
and we obtain a multivariate Gaussian integral. The result of this integral is 
\begin{equation}
 \pp{2\pi}^{N}\sqrt{\det\pp{B}}\left. \bb{\frac{\partial}{\partial \boldsymbol{J}}+\boldsymbol{\chi}_E}^T\bb{\frac{\partial}{\partial \boldsymbol{J}}+\boldsymbol{\chi}_E} \exp\pp{\frac{1}{2}\boldsymbol{J}^T B
 \boldsymbol{J}} \right|_{\boldsymbol{J}=\boldsymbol{0}},
\end{equation}
and we can reduce it to 
\begin{equation}\label{eq:final_number}
 \pp{2\pi}^{N}\sqrt{\det\pp{B}} \pp{ \boldsymbol{\chi}_E^T\boldsymbol{\chi}_E + \text{Tr}\bb{B} },
\end{equation}
which is the contribution to energy from one Gaussian peak of the GKP Wigner function. Putting everything together and doing simplifications on prefactors, we arrive at the result that the average number of excitations in a GKP state described by the generator matrix $S$ and envelope $\pp{\Sigma_E,\boldsymbol{\mu}_E} = \pp{M_E\Sigma_0 M_E^T,\boldsymbol{\mu}_E}$ is
\begin{equation}
  \begin{split}
 \left<\hat{\boldsymbol{\xi}}^T\hat{\boldsymbol{\xi}}\right> = & \pi^{2N}\sqrt{\det\pp{I+\tanh\pp{\Sigma_0^{-1}}}} \sum_{\boldsymbol{\chi}\in\mathcal{M}} c_{\boldsymbol{\chi}}\pp{ \boldsymbol{\chi}_E^T\boldsymbol{\chi}_E + \text{Tr}\bb{M_E \tanh\pp{\Sigma_0^{-1}}M_E^T} } \\ &  \times \exp\pp{\frac{-1}{2}\bb{\boldsymbol{\chi}-\boldsymbol{\mu}_E}^T \bb{\frac{M_E \coth\pp{\Sigma_0^{-1}}M_E^T}{4\pi}}^{-1} \bb{\boldsymbol{\chi}-\boldsymbol{\mu}_E}}.
  \end{split}
\end{equation}
As a final remark, we can also obtain the number of excitations in a sub-system (in a single mode, for example) by restricting the scalar product and trace in \cref{eq:final_number} to be only over elements of this sub-system and rescaling the prefactors accordingly.

\section{Application of \cref{alg:gate_opt}}\label{app:algo1_ex}
In this section, we work out the optimization process of \cref{alg:gate_opt} for the toy logical Clifford circuit on one qubit:
\begin{align}\label{P4_circ}
\begin{quantikz}
\Sigma_0 = I, \mu_0=0\text{ } & \gate{\overline{P}}  & \gate{\overline{P}}  & \gate{\overline{P}}  & \gate{\overline{P}}  &,
\end{quantikz}
\end{align}
applied on a square GKP qubit. This circuit, although trivial in practice, is ideal for demonstrating the mechanics of the algorithm. As inputs of \cref{alg:gate_opt}, we use the parameters $\Sigma_0 = I$ and $\mu_0=0$, as they simplify the computations and make the demonstration more intuitive. For those parameters, both metrics (\cref{eq:met_mu} and \cref{eq:dis_met}) are zero.

The first task of the compiler is to identify all potential implementations of the targeted circuit. For the optimization of the circuit in (\ref{P4_circ}), only the stabilizer group generated by two elements $\mathcal{S}=\left< Z^2, P^2 Z\right> $ is useful, where we used the definitions of \cref{squareGSG} for the operations. At first order of the generators, this group contains the elements
\begin{equation}
  \mathcal{S}_1 = \left\{ I, Z^2, Z^{-2}, P^2 Z, P^2 Z^{-1}, P^{-2} Z, P^{-2} Z^{-1}, P^2Z^3,P^{-2}Z^{-3} \right\}, 
\end{equation}
where every element of the set is the product of the generators at a power of either $-1,0$ or $1$. By multiplying the set $\mathcal{S}_1$ by a representative of $\overline{P}$ (i.e. $P$), we then obtain the different implementations of the logical gate $\overline{P}$ at first order
\begin{equation}
  P\mathcal{S}_1 = \left\{ P, PZ^2, PZ^{-2}, P^3 Z, P^3 Z^{-1}, P^{-1} Z, P^{-1} Z^{-1}, P^3Z^3,P^{-1}Z^{-3} \right\}.
\end{equation}
For the purpose of this demonstration, we retained only the elements with unit magnitude power for all operations
\begin{equation}
  \left\{ P, P^{-1} Z, P^{-1} Z^{-1} \right\}.
\end{equation}
The result of the optimization is not affected by the simplification in this specific case. The second task is to identify the parameters $(M,\boldsymbol{\lambda})$ of each implementation following the form $U = \hat{T}\pp{\boldsymbol{\lambda}}U_M$ described in \cref{alg:gate_opt}. Based on the derivations of \cref{squareGSG}, we find the parametrization
\begin{equation}\label{eq:param_gate_op}
  \left\{ \pp{\begin{bmatrix}1 & 0 \\-1 & 1 \end{bmatrix},\begin{bmatrix}0 & 0 \end{bmatrix}},
  \pp{\begin{bmatrix}1 & 0 \\1 & 1 \end{bmatrix},\begin{bmatrix}0 & 1/\sqrt{2} \end{bmatrix}},
  \pp{\begin{bmatrix}1 & 0 \\1 & 1 \end{bmatrix},\begin{bmatrix}0 & -1/\sqrt{2} \end{bmatrix}}\right\}
\end{equation}
for the operations $\left\{ P, P^{-1} Z, P^{-1} Z^{-1} \right\}$, respectively. Before applying the algorithm, we observe that the set of matrices 
\begin{equation}
  M\pp{n} = \begin{bmatrix}1 & 0 \\-n & 1 \end{bmatrix}
\end{equation}
form a group for which 
\begin{equation}
  M\pp{n}M\pp{m} = M\pp{n+m} \qquad\text{and} \qquad M\pp{n}M\pp{n}^T = \begin{bmatrix}1 & -n \\-n & n^2+1 \end{bmatrix}.
\end{equation}
When we compute the metric \cref{eq:dis_met} for this set of matrices, we get 
\begin{equation}\label{eq:reduc_sq_met}
  \text{Tr}\bb{ \ln^2\pp{M\pp{n}M\pp{n}^T}} = \ln^2\pp{\frac{n^2+2-n\sqrt{n^2+4}}{2}}+\ln^2\pp{\frac{n^2+2+n\sqrt{n^2+4}}{2}}, 
\end{equation}
which is symmetric in $n$ and increasing with the magnitude of $n$. This is useful because the combination of elements of \cref{eq:param_gate_op} is all part of this set. With all the tools in hand, we can now complete the application of \cref{alg:gate_opt}.\\

\noindent In the first round of optimization, we try to optimize the first gate in the logical circuit (\ref{P4_circ}). Initially, the envelope parameters are $\Sigma_0 = I$ and $\boldsymbol{\mu}_0 = \boldsymbol{0}$. At the squeezing level, all implementations in \cref{eq:param_gate_op} are equivalent (the squeezing metric of \cref{eq:reduc_sq_met} is the same for $n=1$ and $n=-1$). Therefore, we turn to the displacement metric to make the discrimination. The second and the third operations both contain displacement, while the first does not. It means that the first implementation of $\overline{P}$ by $P$ is optimal and that after the gate, the envelope parameters are
\begin{equation}
  \Sigma = \begin{bmatrix}1 & -1 \\-1 & 2 \end{bmatrix}\qquad \text{and}\qquad \boldsymbol{\mu}=\boldsymbol{0}.
\end{equation}
We move on to the second gate in circuit (\ref{P4_circ}). We have the choice between implementing the squeezing part of the gate with $M\pp{1}$ or $M\pp{-1}$. While the options are the same as for the first implementation, the optimal choice might be different because the input state is different. Choosing $M\pp{1}$ for the second implementation lead to the metric of \cref{eq:reduc_sq_met} with $n=2$, while choosing $M\pp{-1}$ reverse the squeezing action of the first implemented gate and lead to the same metric with $n=0$. Therefore, the choice $M\pp{-1}$ is optimal. Both options with $M\pp{-1}$ in \cref{eq:param_gate_op} have the same cost for the displacement metric, which means that we can choose at random between the two (we take $\overline{P}=P^{-1}Z$). Following the same rules for the third and fourth implementations, we find that the optimized implementation of circuit (\ref{P4_circ}) is 
\begin{align}\label{P4_circ_imp}
\begin{quantikz}
\Sigma_0 = I, \mu_0=0\text{ } & \gate{P}  & \gate{P^{-1}Z}  & \gate{P^{-1}Z^{-1}}  & \gate{P}  &.
\end{quantikz}
\end{align}
The output of the \cref{alg:gate_opt} for the chosen circuit shows that its action is to first remove as much squeezing as possible, and then recenter the envelope. We observe this mechanism between the first and second implementations, where the $P$ operation is reversed and replaced with an equivalent $Z$ operation. The same logic is applied to the third gate implementation, where the addition of squeezing is necessary, but where we can also remove a unit of displacement in the $Z$ direction. In the general case, the computation is more tedious, and it is difficult to apply \cref{alg:gate_opt} analytically. However, the principles remain the same.

\end{document}